\documentclass{article}

\usepackage[preprint]{neurips_2026}

\usepackage[utf8]{inputenc}
\usepackage[T1]{fontenc}
\usepackage{lmodern}
\usepackage{amsmath, amssymb, amsthm}
\usepackage{graphicx}
\usepackage{setspace}
\usepackage{geometry}
\usepackage{dsfont}  % For indicator function \mathds{1}
\usepackage{bbm}% For indicator function \mathds{1}
\usepackage{amsfonts}       % blackboard math symbols
\usepackage{algorithm}
\usepackage{algorithmic}
\usepackage{amsmath}
\usepackage{amssymb}
\usepackage{xcolor}
\usepackage{tabularx}
\usepackage{booktabs}

\usepackage{hyperref}       % hyperlinks

\usepackage{amsmath, amssymb, amsthm, bm}
\usepackage{mathrsfs}
\usepackage{bbm}

\theoremstyle{definition}

\newtheorem{assumption}{Assumption}

\theoremstyle{remark}

\newcommand\numberthis{\addtocounter{equation}{1}\tag{\theequation}}

\workshoptitle{MLxOR: Mathematical
Foundations and Operational Integration of Machine Learning for Uncertainty-Aware Decision-Making} 

\title{Multifidelity Computer Model Emulation  Via     Diffusion Model Steering and Targeted Maximum Likelihood}

\author{%
  Jongmin Mun
  \\
  University of Southern California
  \\
  \texttt{jongmin.mun@marshall.usc.edu}
}

\begin{document}

\maketitle

\begin{abstract}
We develop a multifidelity method for fusing low-resolution simulations with computationally expensive high-resolution simulations, which are run infrequently and are therefore prone to bias. We formulate this fusion as a constrained optimization under missing-not-at-random (MNAR) selection bias. This formulation searches for the exponentially tilted high-resolution distribution that minimizes KL divergence from the biased baseline, subject to  moment constraints derived from low-resolution simulations.
This optimization requires first estimating the biased baseline conditional density $f$ as a nuisance parameter. We estimate $f$ using a   score-based  diffusion model. To eliminate the generative model's   regularization bias that harms the downstream task, we apply targeted maximum likelihood estimation (TMLE). TMLE debiases $\hat{f}$ via a targeted exponential tilting, rendering the target parameters insensitive to first-order nuisance estimation errors. 
To execute this computationally, we adapt generative model steering, a technique originally developed for human-preference alignment. Using
Feynman-Kac steering with a reward function based on our formulation, we   simultaneously execute the exponential tilts for MNAR and TMLE at inference time, avoiding expensive retraining costs. Code available \href{https://github.com/Jong-Min-Moon/multifidel_emul_by_FK}{here}.
\end{abstract}

\section{Introduction}\label{section:intro}
Computer models are widely used to simulate complex physical processes. For instance, insurers model regional catastrophe risk using proprietary simulators such as AIR   and RMS \citep{ingelsStateArtFuture2024, bertsimas_catastrophe_2024}. Due to subscription and computational cost, high-resolution simulations are run infrequently, meaning available result is often outdated and potentially biased. This gap drives demand for statistical emulators: fast, probabilistic approximations of the underlying model. Combining low-resolution and high-resolution simulation outputs to construct such an emulator is known as multifidelity methods \citep{maMultifidelityComputerModel2022, peherstorferSurveyMultifidelityMethods2018}.
We propose a multifidelity   method that synthesizes techniques from two distinct fields: data fusion via exponential tilting and bias correction from the semiparametric statistical inference literature \citep{guan_data_2026}, and score-based   diffusion \citep{ho_classifierfree_2021} and human-preference alignment \citep{singhalGeneralFrameworkInferencetime2025}  from the   generative modeling literature. 

We formulate emulation as a constrained optimization problem over the distribution space under missing-not-at-random (MNAR) selection bias. Given a biased high-resolution simulation distribution $F$, we model the MNAR mechanism as exponential tilting; We assume that the unbiased high-resolution simulation distribution   is an exponential tilt of $F$. Among the exponentially tilted distributions, we seek the one that minimizes the KL divergence from $F$, subject to moment constraints derived from   unbiased low-resolution simulations. 

The conditional density of $F$, denoted $f$, serves as an infinite-dimensional nuisance parameter within this optimization. We initially estimate $f$ using a diffusion model, building on recent literature that leverages generative models for computer model calibration \citep{Cho03042025} and physical processes \citep{namgungSexDifferencesAutism2024, spitznagelPhysicsGenCanGenerative2025}. Since diffusion models   prioritize generative fidelity over our downstream optimization,
we
 calibrate   this initial   estimate,
 in the spirit of     predict-then-optimize paradigm \citep{elmachtoubSmartPredictThen2022}.
 We achieve this by adopting targeted maximum likelihood estimation (TMLE; \citealp{liTargetedMaximumLikelihood2025}) from semiparametric statistical inference literature, which debiases the machine-learned density $\hat{f}$ via a secondary exponential tilt, rendering the estimation target locally insensitive to first-order nuisance estimation errors.
With the debiased $f$, we then numerically solve the constrained optimization problem over the tilted distribution space. 

We execute both the TMLE and MNAR exponential tilts   at inference time by adapting Feynman-Kac (FK) steering \citep{singhalGeneralFrameworkInferencetime2025} from the generative model alignment literature. By formulating these  tilts as   reward functions, FK steering enables direct sampling from the tilted distributions,   bypassing     the diffusion model retraining.

\section{Problem Formulation}

Let $(X_i, Y_i, R_i) \in \mathcal{X} \times \mathcal{Y} \times \{0, 1\}$ denote the input covariates, continuous outcome, and bias indicator for simulation unit $i \in \{1, \dots, n\}$, governed by the joint distribution $P$. The true marginal input distribution $P_X$ is known (e.g., via experimental design). Our objective is to sample $Y$ from the true conditional distribution $P_{Y|X}$ by fusing two distinct datasets, each with complementary strengths and limitations.

\noindent
\textbf{High-Resolution Misspecified Simulation.}
The first dataset matches the granular resolution of the target distribution $P$ but suffers from model misspecification. We frame this discrepancy as missing-not-at-random (MNAR) selection bias, where   $R$ can depend on both   $X$ and   $Y$. The misspecified high-resolution simulation yields samples $(X_j, Y_j)$ drawn from the biased conditional distribution $F$, defined as $(X, Y) \mid R = 1$. To define the search space for our optimization, we impose a finite-dimensional structural assumption on $F$:

\begin{assumption}[Exponential Tilting; Assumptions 1 and 2 of \citealp{guan_data_2026}]\label{assumption:tilting}
For a known sufficient statistic $\eta : \mathcal{X} \times \mathcal{Y} \to \mathbb{R}$, there exists a parameter $\theta \in \mathbb{R}$ such that:
\begin{equation*}
dP_{Y \mid X=x}(y) \propto dF_{Y \mid X=x}(y) \exp
\bigl(
\theta \eta(x, y)
\bigr) \quad \text{and} \quad \sup_{x \in \mathcal{X}} dF_X(x)/dP_X(x) < \infty.
\end{equation*}
\end{assumption}

\noindent
\textbf{Low-Resolution Correct Simulation.}
We constrain this search space using correctly specified, low-resolution simulations. Given covariate space partitions $A_k \subset \mathcal{X}$, we define   $\gamma_k(x, y) := y \cdot \mathbb{I}(x \in A_k)$. We assume access to the exact population-level moment conditions:
\begin{equation}\label{eq:moment_condition}
\mathbb{E}_P [\gamma_k(X, Y)] = \bar{\gamma}_k, \quad k = 1, \ldots, K.
\end{equation}

\subsection{The Constrained Optimization Problem}

Let $r(x) = dP_X(x)/dF_X(x)$ denote the known covariate density ratio (e.g. from experimental design). Under Assumption \ref{assumption:tilting}, let $Q(\theta, F)$ denote the exponentially tilted distribution:
\begin{equation}\label{tilting:Q}
dQ_{Y\mid X}(\theta, F) \propto dF_{Y\mid X} \cdot \exp
\bigl(
\theta \eta(x, y)
\bigr).
\end{equation}

We seek the parameter $\theta \in \mathbb{R}$ that minimizes the distributional distance between the tilted distribution $Q(\theta, F)$ and the biased baseline $F$, subject to the low-resolution moment constraints \eqref{eq:moment_condition}. Using a change-of-measure formulation, the   optimization is formulated as:
\begin{equation}
\label{constrained_optimization}
\min_{\theta \in \mathbb{R}} \left\{ \mathbb{E}_F \left[ r(X) \cdot \text{KL}(Q_{Y\vert X}(\theta, F) \parallel F_{Y \vert X} ) \right] : \mathbb{E}_F \left[ r(X) \cdot \mathbb{E}_{Q(\theta, F)}[\gamma_k(X, Y) \vert X] \right] = \bar{\gamma}_k \right\}.
\end{equation}
Let $\lambda$ denote the Lagrange multiplier, and define our estimation target as the parameter tuple $\nu = (\theta, \lambda)$. The first-order conditions of  \eqref{constrained_optimization} yield the following estimating equations:
\begin{align*}
\operatorname{DL}(\nu; F) &:= \mathbb{E}_P \left[ \operatorname{Cov}_{Q(\theta,F)}[\eta(X,Y)\mid X]\theta + \operatorname{Cov}_{Q(\theta,F)}[\eta(X,Y),\gamma_k(X,Y)\mid X]\lambda \right] \numberthis \label{estimating_equation_1} \\
\operatorname{M}(\nu; F) &:= \mathbb{E}_P\left[ \mathbb{E}_{Q(\theta,F)}[\gamma_k(X,Y)\mid X] \right] - \bar{\gamma}_k. \numberthis \label{estimating_equation_2}
\end{align*}

The moment constraints under $Q(\theta, F)$ induce non-convexity and 
their evaluation
requires estimation of   conditional density of $F_{Y|X}$, denoted $f(y|x)$     (details in Appendix \ref{section:why_f}). Consequently, we focus on identifying a valid local root rather than a global optimum, with   nuisance parameter $f$.

\section{Proposed Method: Diffusion, Debiasing, and Optimization, with Steering}
To solve the optimization problem \eqref{constrained_optimization}, we   first estimate the biased baseline conditional density, $f(y \mid x)$, and then sample from its exponentially tilted version. We achieve this through  three stages.

\subsection{Conditional Density Estimation via Diffusion}
We estimate the baseline density $f$ from the misspecified high-resolution data using a continuous-time score-based conditional diffusion model with classifier-free guidance \citep{ho_classifierfree_2021}. Because the outcome $Y$ is a one-dimensional scalar, we use a lightweight, conditioned residual MLP architecture rather than standard high-dimensional image networks (details in Appendix \ref{section:diffusion_architecture}). 

\subsection{Debiasing via Targeted Maximum Likelihood Estimation (TMLE)}
Because our end-goal target is the parameter tuple $\nu$, the conditional density $f$ acts as a nuisance parameter. Diffusion models are  optimized for generalizability and smooth generation, making the initial density estimate $\hat{f}$ highly susceptible to regularization bias. If left uncorrected, this bias propagates into the estimation of $\nu$.
To mitigate this, we update the initial diffusion-based estimate via a targeted exponential fluctuation (model derived in Appendix \ref{section:TMLE_submodel}):
\begin{equation} \label{eq:tmle_submodel}
f_\epsilon(y \mid x) \propto f(y \mid x) \exp
\bigl(
\epsilon^\top D^\ast (x, y)
\bigr) 
\end{equation}

Here, $\epsilon \in \mathbb{R}^{K+1}$ is the fluctuation parameter, and $D^\ast (x, y) \in \mathbb{R}^{K+1}$ is the canonical gradient associated with our estimating equations (derived in Appendix \ref{section:derive_naive_estimating_equation} and \ref{section:proof:theorem:canonical_gradient}; numerical evaluation in Appendix \ref{section:can_grad_eval}). Geometrically, $D^\ast$ quantifies the first-order effect that an infinitesimal perturbation in the data-generating distribution has on the target parameter $\nu$. We optimize $\epsilon$ via maximum likelihood until the empirical mean of the estimated canonical gradient is driven to zero. This ensures the effect of first-order estimation error of $f$ on $\nu$ vanishes, leaving only negligible second-order impacts on $\nu$ \citep{liTargetedMaximumLikelihood2025}.

\subsection{Execution via Feynman-Kac (FK) Steering.}
Both the MNAR   modeling and the TMLE regularization correction rely on exponential tilt of the baseline conditional density $f$ (using $\theta \eta(x, y)$ and $\epsilon^\top D^\ast (x, y)$, respectively). Retraining a generative model for tilting is computationally expensive \citep{ICLR2025_852f5096}. 
Instead, we treat these tilting terms as reward functions in generative model steering for human preference alignment and execute the tilts  at inference time. Specifically, we use Feynman-Kac (FK) steering \citep{singhalGeneralFrameworkInferencetime2025}, which simulates $M$ concurrent diffusion trajectories (particles). Instead of applying the exponential tilt in only one step of generation, it defines a sequence of intermediate potential functions along the reverse diffusion trajectory. As generation unfolds, particles are mutated by the base diffusion score, continuously evaluated against these intermediate potentials, and   resampled in high-reward regions. This   yields samples from the tilted distributions without     retraining. Our implementation detail is provided in Appendix \ref{section:fk}. 

\subsection{End-to-End Workflow}
Our complete estimation pipeline integrates the core components (diffusion, TMLE, KL optimization, FK steering) into a unified procedure consisting of three primary phases:
\begin{enumerate}
    \item \textbf{Base Density Estimation:} We first estimate the baseline conditional density $f$ from the biased high-resolution simulation data using a continuous-time conditional diffusion model. Full architectural details for this score-based estimator are provided in Appendix \ref{section:diffusion_architecture}.
    
    \item \textbf{TMLE Debiasing:} Next, we correct the diffusion model's  regularization bias by applying targeted maximum likelihood estimation (TMLE). This targeted exponential fluctuation is executed efficiently via inference-time Feynman-Kac (FK) steering, as detailed in Appendix \ref{section:solve_mle} and   Algorithm \ref{alg:tmle_grad_ascent}.
    
    \item \textbf{Constrained Optimization:} Finally, using the debiased density, we numerically solve the core constrained KL optimization problem \eqref{constrained_optimization}, using 
    mini-batch stochastic gradient
descent-ascent (SGDA) procedure.
    To bypass the  computational cost of iteratively retraining the generative model, we once again leverage FK steering to instantiate the MNAR exponential tilts and evaluate the necessary expectations directly at inference time (detailed in Appendix \ref{section:solve_estimating_eq}).
\end{enumerate}
\section{Numerical Results}
 
Table \ref{tab:experiments_comparison} empirically demonstrates that our proposed framework   outperforms the uncalibrated high-resolution simulator across three   physical modeling scenarios, especially when TMLE was applied (full experimental details are provided in Appendix \ref{section:numerical}).
\begin{table}[b!]
\centering
\small
\caption{Three experimental setups and RMSE results.}
\label{tab:experiments_comparison}
\begin{tabular}{@{} >{\raggedright\arraybackslash}p{3.2cm} >{\raggedright\arraybackslash}p{3.1cm} >{\raggedright\arraybackslash}p{3.1cm} >{\raggedright\arraybackslash}p{3.1cm} @{}}
\toprule
& \textbf{Gas Compressibility} & \textbf{Linear Compartment} & \textbf{Surface Adsorption} \\
\midrule
\textbf{$Y$} & Volumetric response & Terminal compartment state & Adsorbed volume \\
\addlinespace
\textbf{True Model} & Nonlinear compressibility curve & ODE with nonlinear equilibrium $\mu(X)$ & Langmuir isotherm (physical saturation) \\
\addlinespace
\textbf{Biased Model} & Linear ideal-gas law & ODE with linear $\mu(X)$ and fast rate $\gamma$ & Henry's Law (linear) \\
\addlinespace
\textbf{Bias Source} & Omits nonlinear compressibility bend & Misspecified equilibrium target and kinetic rate & Omits asymptotic saturation; calibration shift \\
\addlinespace
\textbf{RMSE (original)} & 5.05 & 6.74 & 8.37 \\
\addlinespace
\textbf{RMSE (our method without TMLE)} & 3.63 & 5.54 & 5.66 \\
\addlinespace
\textbf{RMSE (our method with TMLE)} & \textbf{3.50} & \textbf{5.28} & \textbf{5.47} \\
\bottomrule
\end{tabular}
\end{table}
 
 \section{Future Direction}
Future research will expand on both theoretical and algorithmic fronts. On the theoretical side, we aim to establish the asymptotic consistency and asymptotic normality of our proposed estimator, which will enable   uncertainty quantification and formal confidence intervals for the fused emulator. 
Algorithmically, we plan to replace the current   SGDA optimization procedure with a one-step estimator \citep{laan_targeted_2011}, which  refines the initial   estimate via a targeted Newton-Raphson step. One-step estimator offers stronger theoretical guarantees for   optimality and statistical convergence rate.
Additionally, we will incorporate cross-fitting to separate the estimation of $f$ from the downstream optimization. This sample-splitting approach enables more effective bias correction and bypasses the Donsker condition, allowing us to establish rigorous convergence guarantees.

\bibliographystyle{apalike}
\bibliography{reference}

\appendix

\section{Derivation of the Estimating Equations}\label{section:derive_naive_estimating_equation}
The first-order conditions of the core optimization problem \eqref{constrained_optimization} yield the following estimating equations:
\begin{align*}
\operatorname{DL}(\nu; F) &:= \mathbb{E}_P \left[ \operatorname{Cov}_{Q(\theta,F)}[\eta(X,Y)\mid X]\theta + \operatorname{Cov}_{Q(\theta,F)}[\eta(X,Y),\gamma_k(X,Y)\mid X]\lambda \right]  \\
\operatorname{M}(\nu; F) &:= \mathbb{E}_P\left[ \mathbb{E}_{Q(\theta,F)}[\gamma_k(X,Y)\mid X] \right] - \bar{\gamma}_k. 
\end{align*}
This population estimating equations is identical to the binary outcome case of \cite{guan_data_2026}.
Because \eqref{constrained_optimization} is a constrained optimization problem, we introduce the Lagrangian by augmenting the objective function with the constraint multiplied by the vector of Lagrange multipliers $\lambda \in \mathbb{R}^K$. Taking the partial derivatives of the Lagrangian with respect to the optimization variable $\theta \in \mathbb{R}$ and the dual parameter $\lambda$ yields the   population estimating equations.
The derivation relies critically on the exponential tilting representation in Assumption~\ref{assumption:tilting} together with the KL divergence objective. The derivation, which is omitted in \cite{guan_data_2026}, is provided here for completeness.

In semiparametric problems, the parameter of interest is typically expressed as an explicit functional of the underlying data-generating distribution. In our setting, however, the parameter of interest, denoted by $\nu = (\theta, \lambda)$, is defined   implicitly as the solution to the system of estimating equations
\[
\Psi(\nu, f)
=
\begin{bmatrix}
\mathrm{DL}(\nu; F) \\
M(\nu; F)
\end{bmatrix}
=
\mathbf{0}.
\]

Now we derive the population estimation equations \eqref{estimating_equation_2}.
The optimization formulation uses a change of measure trick. Notice the term $r(X) = dP_X(x)/dF_X(x)$. If we multiply an expectation over the biased high-resolution distribution $F$ by $r(X)$, it mathematically transforms it into an expectation over the true population distribution $P$. Recognizing this, our optimization problem is equivalently formulated as:
\begin{equation}
\min_{\theta} \mathbb{E}_P \left[ \text{KL}(Q_{Y \mid X}(\theta, F) \parallel F_{Y \mid X}(F)) \right] \quad \text{subject to} \quad \mathbb{E}_P \left[ \mathbb{E}_{Q(\theta, F)}[\gamma_k(X,Y) \mid X] \right] = \bar{\gamma}_k
\end{equation}

We combine the objective function and the constraint into a single Lagrangian by introducing a Lagrange multiplier, $\lambda$, which penalizes the objective if the moment constraint is violated:
\begin{equation}
\mathcal{L}(\theta, \lambda) = \mathbb{E}_P \left[ \text{KL}(Q_{Y \mid X}(\theta, F) \parallel F_{Y \mid X}(F)) \right] + \lambda^T \left( \mathbb{E}_P \left[ \mathbb{E}_{Q(\theta, F)}[\gamma_k(X, Y) \mid X] \right] - \bar{\gamma}_k \right)
\end{equation}

We seek the stationary point where the derivatives of $\mathcal{L}$ with respect to $\theta$ and $\lambda$ are exactly zero.

\subsection{Partial Derivative by Tilting Parameter}
To derive $\mathrm{DL}(\nu; F)$,
we take the derivative of $\mathcal{L}$ with respect to $\theta$. Assuming regularity conditions allow us to pass the derivative inside the expectation via the Dominated Convergence Theorem, we analyze the terms inside $\mathbb{E}_P$.

\subsubsection{The KL Divergence Term}
By definition, $\text{KL}(Q \parallel S) = \mathbb{E}_Q [\log(dQ/dS)]$. Since the ratio $dQ/dS$ is proportional to $\exp(\theta^T \eta)$, we have:
\begin{equation}
\log\left(\frac{dQ}{dS}\right) = \theta^T \eta(X,Y) - A(X, \theta)
\end{equation}
Here, $A(X, \theta)$ is the log-partition function (the normalizing constant ensuring the tilted density integrates to 1 over the continuous space of $Y$). Taking the expected value under $Q$ yields:
\begin{equation}
\text{KL}(Q \parallel S) = \theta^T \mathbb{E}_Q[\eta(X,Y) \mid X] - A(X, \theta)
\end{equation}
Note that $A(X, \theta)$ is a function of $X$ and $\theta$, but unaffected by the expectation over $Y$. We take the gradient with respect to $\theta$ using the product rule on the first term:
\begin{equation}
\nabla_\theta \text{KL} = \mathbb{E}_Q[\eta(X,Y) \mid X] + \left(\nabla_\theta \mathbb{E}_Q[\eta(X,Y) \mid X]\right) \theta - \nabla_\theta A(X, \theta)
\end{equation}
A fundamental property of exponential families is that the derivative of the log-partition function is exactly the expected value of the sufficient statistic. Therefore, $\nabla_\theta A(X, \theta) = \mathbb{E}_Q[\eta(X,Y) \mid X]$. This mathematically cancels out the first and third terms, leaving:
\begin{equation}
\nabla_\theta \text{KL} = \left(\nabla_\theta \mathbb{E}_Q[\eta(X,Y) \mid X]\right) \theta
\end{equation}
A second key property of exponential families is that the derivative of the expectation of the sufficient statistic yields its covariance matrix. Therefore, $\nabla_\theta \mathbb{E}_Q[\eta(X,Y) \mid X] = \text{Cov}_Q[\eta(X,Y) \mid X]$. This gives us the first half of the derivative:
\begin{equation}
\nabla_\theta \text{KL} = \text{Cov}_Q[\eta(X,Y) \mid X] \theta
\end{equation}

\subsubsection{The Constraint Term}
Next, we take the gradient of the constraint term, $\lambda^T \mathbb{E}_Q[\gamma_k(X,Y) \mid X]$, with respect to $\theta$. Applying the chain rule and reusing the property of exponential families, the derivative of the expectation of any function $\gamma$ under an exponential tilt is the covariance between that function and the sufficient statistic $\eta$:
\begin{equation}
\nabla_\theta \left( \lambda^T \mathbb{E}_Q[\gamma_k(X,Y) \mid X] \right) = \text{Cov}_Q[\eta(X,Y), \gamma_k(X,Y) \mid X] \lambda
\end{equation}

\subsubsection{Conclusion}
Combining these two derivatives and taking the outer expectation over $P$, we obtain the exact formula for $\mathrm{DL}(\nu; F)$ \eqref{estimating_equation_1}:
\begin{equation}
\nabla_\theta \mathcal{L} = \mathbb{E}_P \left[ \text{Cov}_{Q(\theta, F)} [\eta(X,Y) \mid X] \theta + \text{Cov}_{Q(\theta, F)} [\eta(X,Y), \gamma_k(X,Y) \mid X] \lambda \right] = \mathrm{DL}(\nu; F)
\end{equation}

\subsection{Partial Derivative by Lagrangian Multiplier}
To derive $M(\nu; F)$,
we take the derivative of the Lagrangian with respect to the Lagrange multiplier $\lambda$ simply returns the constraint evaluated at the current parameters:
\begin{equation}
\nabla_\lambda \mathcal{L} = \mathbb{E}_P \left[ \mathbb{E}_{Q(\theta, F)}[\gamma_k(X, Y) \mid X] \right] - \bar{\gamma}_k = M(\nu; F)
\end{equation}
This  matches the moment condition 
\eqref{estimating_equation_2}
used in the population estimating equations.

\section{Why Continuous Outcomes Require Conditional Density Estimation}\label{section:why_f}
Consider a simple     linear shift, $\eta(x, y) = y$. Under these assumptions, the estimating equation becomes:
\begin{equation}
\mathbb{E}_F \left[ r(X) \cdot \mathbb{I}(X \in A_k) \cdot \mathbb{E}_{Q(\theta, F)}[Y \mid X] \right] - \bar{\gamma}_{P, k} = 0.
\end{equation}
The inner conditional expectation expands into a ratio of two integrals over the baseline conditional density $F(y \mid x)$:
\begin{equation}
\mathbb{E}_{Q(\theta, F)}[Y \mid X=x] = \frac{\int y \cdot F(y \mid x) \exp(\theta y) \, dy}{\int F(y \mid x) \exp(\theta y) \, dy}.
\end{equation}
Notice that the denominator is exactly the conditional Moment Generating Function (MGF) of $F(y \mid x)$ evaluated at $\theta$. This mathematical structure reveals two critical properties of our problem:

\begin{enumerate}
    \item \textbf{Non-Convexity:} Because the optimization variable $\theta$ resides within the exponential function of both the numerator and the denominator, the expectation is highly nonlinear with respect to $\theta$. This nonlinearity forces the feasible region of the constrained optimization to be non-convex. Consequently, simply finding a root for equations \eqref{estimating_equation_1} and \eqref{estimating_equation_2} using standard solvers does not guarantee convergence to a global optimum.
    \item \textbf{Full Density Estimation Requirement:} Evaluating the MGF  depends on the higher-order moments and tail behavior of the underlying data. Therefore, the full conditional density must be estimated as the nuisance parameter. This is different from the binary outcome setting of \citet{guan_data_2026}, where the nuisance parameter could be collapsed into a   conditional outcome expectation  rather than requiring the entire continuous conditional distribution.
\end{enumerate}

\section{Derivation of Canonical Gradient}\label{section:proof:theorem:canonical_gradient}
 
Let $\nu = (\theta, \lambda)$ be the finite-dimensional parameter defined implicitly by the estimating equations $\Psi(\nu, f) = [\mathrm{DL}(\nu; F)^\top, M(\nu; F)^\top]^\top = \mathbf{0}$, where $f(y \mid x)$ is the infinite-dimensional nuisance conditional density. 

Let $f_\epsilon(y \mid x) = f(y \mid x)(1 + \epsilon h(y \mid x))$ denote a regular parametric submodel with a valid, conditionally mean-zero perturbation direction $h(y \mid x)$. The canonical gradient of $\Psi(\nu, f)$ with respect to $f$ under the baseline distribution $F$ is the unique joint vector function $D^\ast(X, Y) = [\mathrm{D}_{\mathrm{DL}}(X, Y)^\top, \mathrm{D}_M(X, Y)^\top]^\top$ that satisfies the G\^ateaux derivative mapping:
\[
\left. \frac{\partial}{\partial \epsilon} \Psi(\nu; f_\epsilon) \right|_{\epsilon=0} = \mathbb{E}_F \left[ D^\ast(X, Y) \, h(Y \mid X) \right].
\]
The exact closed-form components of the canonical gradient are given by:
\[
\mathrm{D}_M(X, Y) = r(X) w(X, Y; \theta, f) \Big( \gamma_k(X, Y) - \mathbb{E}_{Q(\theta, F)}[\gamma_k(X, Y) \mid X] \Big)
\]
and
\[
\mathrm{D}_{\mathrm{DL}}(X, Y) = r(X) w(X, Y; \theta, f) \Big( W_0(X, Y) - \mathbb{E}_{Q(\theta, F)}[W_0(X, Y) \mid X] \Big),
\]
where:
\begin{itemize}
    \item $r(X) = dP_X(X) / dF_X(X)$ is the known covariate density ratio.
    \item $w(X, Y; \theta, f) = \exp(\theta \eta(X, Y)) / Z_0(X)$ is the exponential tilting weight, with $Z_0(X) = \int \exp(\theta \eta(X, z)) f(z \mid X) \, dz$.
    \item $Q(\theta, F)$ is the tilted probability measure.
    \item $W_0(X, Y) = \Delta \eta(X, Y) \left( \Delta \eta(X, Y)^\top \theta + \Delta \gamma_k(X, Y)^\top \lambda \right)$, where $\Delta$ denotes the residual of the variable centered by its conditional expectation under $Q(\theta, F)$.
\end{itemize}

\textbf{Remark:} Both canonical gradient components share a unifying geometric structure: $r(X) w(X, Y) \left[ T(X, Y) - \mathbb{E}_Q[T(X, Y) \mid X] \right]$. This structure ensures that $D^\ast(X,Y)$ is conditionally mean-zero under $F$, a property critical for constructing a valid Target Maximum Likelihood Estimation (TMLE) exponential fluctuation submodel. By the Implicit Function Theorem, driving the empirical mean of $D^\ast(X, Y)$ to zero achieves Neyman orthogonality for the target parameter $\nu$.

\subsection{Derivation Strategy}
In many standard semiparametric problems, the parameter of interest is defined explicitly as a functional of the underlying data-generating distribution, denoted by $\psi(P)$. For example, under the standard unconfoundedness assumption, the average treatment effect can be written as
\[
\psi(P)
=
\mathbb{E}_P\!\left[
\mathbb{E}_P[Y \mid X,T=1]
-
\mathbb{E}_P[Y \mid X,T=0]
\right].
\]
Because the target parameter is available in closed form, its pathwise derivative can be obtained directly by taking the Gâteaux derivative of the functional, leading to the efficient influence curve (EIC).

In contrast, the parameter in our problem, $\nu=(\theta,\lambda)$, is not available as an explicit functional of the distribution. Instead, it is defined implicitly as the solution to the estimating equations
\[
\Psi(\nu,f)
=
\begin{bmatrix}
\mathrm{DL}(\nu;f)\\
M(\nu;f)
\end{bmatrix}
=
\mathbf{0}.
\]
Since there is no explicit representation of the form $\nu=g(F)$, the parameter cannot be differentiated directly. Instead, we invoke the infinite-dimensional Implicit Function Theorem to characterize its pathwise derivative.

Let $f_\epsilon$ denote a regular parametric submodel of the nuisance density indexed by $\epsilon$ (this notation overlaps with TMLE fluctuation parameter; in this section, we strictly use $\epsilon$ Gâteaux derivative, not TMLE fluctuation), and let $\nu(\epsilon)$ denote the corresponding solution of the estimating equations. By definition,
\[
\Psi(\nu(\epsilon),f_\epsilon)=\mathbf{0},
\]
for every $\epsilon$ in a neighborhood of zero. Differentiating both sides with respect to $\epsilon$ and applying the multivariate chain rule yields
\[
\frac{\partial \Psi}{\partial \nu}
\frac{\partial \nu}{\partial \epsilon}
+
\frac{\partial \Psi}{\partial \epsilon}
=
\mathbf{0},
\]
where all derivatives are evaluated at $\epsilon=0$. The first term,
\[
\frac{\partial \Psi}{\partial \nu},
\]
is the Jacobian of the estimating equations with respect to the finite-dimensional parameter $\nu$, while
\[
\frac{\partial \Psi}{\partial \epsilon},
\]
is the Gâteaux derivative of the estimating equations with respect to the infinite-dimensional nuisance parameter. 

Assuming that the Jacobian is nonsingular, the Implicit Function Theorem gives
\[
\frac{\partial \nu}{\partial \epsilon}
=
-
\left(
\frac{\partial \Psi}{\partial \nu}
\right)^{-1}
\frac{\partial \Psi}{\partial \epsilon}.
\]
Consequently, if the Gâteaux derivative admits the representation
\[
\frac{\partial \Psi}{\partial \epsilon}
=
\mathbb{E}_F\!\left[
D^\ast(X,Y)\,
h(Y\mid X)
\right],
\]
then the pathwise derivative of the parameter is
\[
\frac{\partial \nu}{\partial \epsilon}
=
\mathbb{E}_F\!\left[
-
\left(
\frac{\partial \Psi}{\partial \nu}
\right)^{-1}
D^\ast(X,Y)\,
h(Y\mid X)
\right].
\]
By definition, the quantity multiplying the score function $h(Y\mid X)$ is the canonical gradient, or efficient influence curve, of the parameter $\nu$. Hence,
\[
\mathrm{EIC}_{\nu}(X,Y)
=
-
\left(
\frac{\partial \Psi}{\partial \nu}
\right)^{-1}
D^\ast(X,Y).
\]

This characterization shows that the efficient influence curve for $\nu$ is simply a linear transformation of $D^\ast(X,Y)$. Therefore, it is not necessary to derive $\mathrm{EIC}_{\nu}$ explicitly. Instead, it suffices to construct a TMLE fluctuation submodel whose score is equal to $D^\ast(X,Y)$. Updating the nuisance estimate until the empirical mean of $D^\ast(X,Y)$ is zero forces the empirical Gâteaux derivative of the estimating equations to vanish. Equivalently, the estimating equations become locally insensitive to first-order perturbations of the nuisance parameter, thereby achieving Neyman orthogonality and eliminating first-order bias arising from nuisance estimation.

\subsection{Gâteaux Derivative Computation for the Second Estimating Equation}
Our goal is to find the function $\mathrm{D}_M(X, Y)$ (the influence curve component) such that for a valid perturbation direction $h$, the directional derivative of $M(\nu; F)$ is equal to the inner product $\mathbb{E}_F[\mathrm{D}_M(X, Y) h(Y\vert{}X)]$.

\subsubsection{Define the Perturbation Path}
For a continuous conditional density $f(y\vert{}x)$ under the underwriting distribution $F$, we define the multiplicative parametric submodel:

$$f_\epsilon(y\vert{}x) = f(y\vert{}x)(1 + \epsilon h(y\vert{}x))$$

To ensure that $f_\epsilon(y\vert{}x)$ remains a valid probability density (i.e., integrates to 1) for small $\epsilon$, the perturbation direction $h(y\vert{}x)$ must be mean-zero under $F$:

$$\mathbb{E}_F [h(Y\vert{}X) \mid X=x] = \int h(y\vert{}x) f(y\vert{}x) dy = 0 \quad \forall x \in \mathcal{X}$$

\subsubsection{Perturb the Exponentially Tilted Measure}\label{step2}

The estimating equation $M(\nu; F)$ is driven by the tilted distribution $Q(\theta, F)$. Its conditional density is:

\begin{equation}\label{q_zero}
q_0(y\vert{}x) = \frac{\exp(\theta \eta(x, y)) f(y\vert{}x)}{Z_0(x)}
\end{equation}

where the normalizing constant is $Z_0(x) = \int \exp(\theta \eta(x, z)) f(z\vert{}x) dz$.

When we plug our perturbed density $f_\epsilon$ into the tilted measure, the new density $q_\epsilon(y\vert{}x)$ has a perturbed normalizing constant:

$$Z_\epsilon(x) = \int \exp(\theta \eta(x, z)) f(z\vert{}x) (1 + \epsilon h(z\vert{}x)) dz$$

Taking the derivative of $Z_\epsilon(x)$ with respect to $\epsilon$ at $\epsilon=0$ gives:

$$\frac{\partial}{\partial \epsilon} Z_\epsilon(x) \bigg\vert{}_{\epsilon=0} = \int \exp(\theta \eta(x, z)) f(z\vert{}x) h(z\vert{}x) dz = Z_0(x) \mathbb{E}_{Q(\theta, F)}[h(Y\vert{}X) \mid X=x]$$

Now, applying the quotient rule to $q_\epsilon(y\vert{}x)$, we find the score of the tilted distribution with respect to $\epsilon$:

\begin{align*}
   & \frac{\partial}{\partial \epsilon} q_\epsilon(y\vert{}x) \bigg\vert{}_{\epsilon=0} =
    \\
    &\frac{\exp(\theta \eta(x, y)) f(y\vert{}x) h(y\vert{}x) Z_0(x) - \exp(\theta \eta(x, y)) f(y\vert{}x) \left( Z_0(x) \mathbb{E}_{Q(\theta, F)}[h(Y\vert{}X) \mid X=x] \right)}{Z_0(x)^2}
\end{align*}

Factoring out $q_0(y\vert{}x)$, this simplifies   to:

$$\frac{\partial}{\partial \epsilon} q_\epsilon(y\vert{}x) \bigg\vert{}_{\epsilon=0} = q_0(y\vert{}x) \left( h(y\vert{}x) - \mathbb{E}_{Q(\theta, F)}[h(Y\vert{}X) \mid X=x] \right)$$

\subsubsection{Differentiate the Estimating Equation}

Recall the moment estimating equation:

$$M(\nu; F) = \mathbb{E}_P \left[ \int \gamma_k(X, Y) q_0(y\vert{}x) dy \right] - \bar{\gamma}_k$$

We plug in $q_\epsilon$ and differentiate under the integral sign:

$$\frac{\partial}{\partial \epsilon} M(\nu; f_\epsilon) \bigg\vert{}_{\epsilon=0} = \mathbb{E}_P \left[ \int \gamma_k(X, Y) \frac{\partial}{\partial \epsilon} q_\epsilon(y\vert{}x) \bigg\vert{}_{\epsilon=0} dy \right]$$

Substitute the score we derived in Appendix \ref{step2}:

$$\frac{\partial}{\partial \epsilon} M(\nu; f_\epsilon) \bigg\vert{}_{\epsilon=0} = \mathbb{E}_P \left[ \int \gamma_k(X, Y) q_0(y\vert{}x) \left( h(y\vert{}x) - \mathbb{E}_{Q(\theta, F)}[h(Y\vert{}X) \mid X] \right) dy \right]$$

Notice that this integral is simply the conditional covariance under the tilted measure $Q$:

$$\frac{\partial}{\partial \epsilon} M(\nu; f_\epsilon) \bigg\vert{}_{\epsilon=0} = \mathbb{E}_P \left[ \text{Cov}_{Q(\theta, F)} (\gamma_k(X, Y), h(Y\vert{}X) \mid X) \right]$$

\subsubsection{Isolate the Direction  and Change the Measure}

To extract the canonical gradient, we need to rewrite this expectation as an inner product with $h(Y\vert{}X)$ under the underwriting distribution $F$.

First, we use a standard property of covariance ($\text{Cov}(A, B) = \mathbb{E}[(A - \mathbb{E}[A])B]$) to rewrite the inner integral:

\begin{align*}
&\int \gamma_k(X, Y) q_0(y\vert{}x) \left( h(y\vert{}x) - \mathbb{E}_{Q(\theta, F)}[h(Y\vert{}X) \mid X=x] \right) dy 
\\& \quad = \int \left( \gamma_k(X, Y) - \mathbb{E}_{Q(\theta, F)}[\gamma_k(X, Y) \mid X=x] \right) q_0(y\vert{}x) h(y\vert{}x) dy
\end{align*}

Next, we expand the outer expectation $\mathbb{E}_P$ as an integral over the population covariate distribution $P_X$. By Assumption 2, we know $dP_X(x) = r(x) dF_X(x)$, where $r(x)$ is the covariate density ratio.

$$= \int \int \left( \gamma_k(X, Y) - \mathbb{E}_{Q(\theta, F)}[\gamma_k(X, Y) \mid X=x] \right) q_0(y\vert{}x) h(y\vert{}x) dy \, r(x) dF_X(x)$$

Finally, we expand $q_0(y\vert{}x) = \frac{\exp(\theta \eta(x, y))}{Z_0(x)} f(y\vert{}x)$. 
Let $w(x, y; \theta, f) = \frac{\exp(\theta \eta(x, y))}{Z_0(x)}$ be the tilting weight. Rearranging the terms gives:

$$= \int \int r(x) w(x, y; \theta, f) \left( \gamma_k(X, Y) - \mathbb{E}_{Q(\theta, F)}[\gamma_k(X, Y) \mid X=x] \right) h(y\vert{}x) f(y\vert{}x) dy \, dF_X(x)$$

Because $f(y\vert{}x) dy \, dF_X(x) = dS(x, y)$, this is exactly an expectation under the underwriting distribution $F$:

$$\frac{\partial}{\partial \epsilon} M(\nu; f_\epsilon) \bigg\vert{}_{\epsilon=0} = \mathbb{E}_F \left[ r(X) w(X, Y; \theta, f) \left( \gamma_k(X, Y) - \mathbb{E}_{Q(\theta, F)}[\gamma_k(X, Y) \mid X] \right) h(Y\vert{}X) \right]$$

\subsubsection{Conclusion: the Canonical Gradient}

By representing the Gâteaux derivative as $\mathbb{E}_F[\mathrm{D}_M(X, Y) h(Y\vert{}X)]$, we have successfully isolated the canonical gradient (the efficient influence curve component) for $M(\nu; F)$ with respect to the nuisance density $f$:

$$\mathrm{D}_M(X, Y) = r(X) w(X, Y; \theta, f) \left( \gamma_k(X, Y) - \mathbb{E}_{Q(\theta, F)}[\gamma_k(X, Y) \mid X] \right)$$

\subsection{Gâteaux Derivative Computation for the First Estimating Equation}
  Because we are taking the Gâteaux derivative of a covariance, we can use a   calculus-of-variations trick that  bypasses the   product-rule expansions used in the   Appendix A.4.2. of \cite{guan_data_2026}.

\subsubsection{Simplify the Notation}
Let us define the centered variables under the tilted measure $Q(\theta, F)$. For any function $g(X, Y)$, let its conditional expectation under $Q$ be $\mu_{g, Q}(X) = \mathbb{E}_{Q(\theta, F)}[g(X, Y) \mid X]$, and its centered version be:

$$\Delta g(X, Y) = g(X, Y) - \mu_{g, Q}(X)$$

Using this notation, we can rewrite the estimating equation as a sum of two conditional covariance matrices acting on the parameter vectors $\theta$ and $\lambda$:

$$\mathrm{DL}(\nu; F) = \mathbb{E}_P \left[ \mathbb{E}_{Q(\theta,F)}[\Delta \eta(X,Y) \Delta \eta(X,Y)^\top \mid X] \theta + \mathbb{E}_{Q(\theta,F)}[\Delta \eta(X,Y) \Delta \gamma_k(X,Y)^\top \mid X] \lambda \right]$$

To make the subsequent integration easier to read, we can pull the vectors $\theta$ and $\lambda$ inside the expectations, defining a single combined function $W(X, Y)$:

$$W(X, Y) = \Delta \eta(X, Y) \left( \Delta \eta(X, Y)^\top \theta + \Delta \gamma_k(X, Y)^\top \lambda \right)$$

This allows us to write the estimating equation simply as:

$$\mathrm{DL}(\nu; F) = \mathbb{E}_P \left[ \mathbb{E}_{Q(\theta,F)}[ W(X, Y) \mid X ] \right] = \mathbb{E}_P \left[ \int W(x,y) q_0(y\vert{}x) dy \right]$$

\subsubsection{The Covariance Derivative Trick}
We need to differentiate $\mathrm{DL}(\nu; f_\epsilon)$ with respect to $\epsilon$ at $\epsilon=0$.
Recall from the $M(\nu; F)$ derivation that the score of the tilted density is $s_Q(y\vert{}x) = h(y\vert{}x) - \mathbb{E}_{Q}[h(Y\vert{}X)\vert{}X]$, so that $\frac{\partial}{\partial \epsilon} q_\epsilon(y\vert{}x) \big\vert{}_{\epsilon=0} = s_Q(y\vert{}x) q_0(y\vert{}x)$.
Notice that $W(X, Y)$ is defined using centered variables $\Delta \eta$ and $\Delta \gamma$, which depend on $\epsilon$ because their conditional expectations $\mu_{\eta, Q_\epsilon}$ and $\mu_{\gamma, Q_\epsilon}$ shift as the density perturbs.

However, when differentiating a covariance, the derivatives of the means vanish at $\epsilon = 0$. Let $C_\epsilon = \mathbb{E}_{Q_\epsilon}[(U - \mu_{u,\epsilon})(V - \mu_{v,\epsilon})]$. Differentiating using the product rule yields:

$$\frac{\partial}{\partial \epsilon} C_\epsilon \bigg\vert{}_{\epsilon=0} = \int \left[ - \mu_{u}' (V - \mu_v) - \mu_{v}' (U - \mu_u) \right] q_0 + \int (U - \mu_u)(V - \mu_v) q_0'$$

Because $\int (V - \mu_v) q_0 = 0$ and $\int (U - \mu_u) q_0 = 0$, the first integral equals zero. Therefore, we only need to differentiate the density $q_\epsilon$, treating the centering terms inside $W(X, Y)$ as constants fixed at $\epsilon=0$:

$$\frac{\partial}{\partial \epsilon} \mathbb{E}_{Q_\epsilon}[W_\epsilon(X, Y) \mid X] \bigg\vert{}_{\epsilon=0} = \int W_0(x, y) s_Q(y\vert{}x) q_0(y\vert{}x) dy$$

\subsubsection{Differentiate the Estimating Equation}

Applying this trick, we differentiate $\mathrm{DL}(\nu; f_\epsilon)$ by passing the derivative through the population expectation $\mathbb{E}_P$:

$$\frac{\partial}{\partial \epsilon} \mathrm{DL}(\nu; f_\epsilon) \bigg\vert{}_{\epsilon=0} = \mathbb{E}_P \left[ \int W_0(x,y) s_Q(y\vert{}x) q_0(y\vert{}x) dy \right]$$

Substitute the score $s_Q(y\vert{}x) = h(y\vert{}x) - \mathbb{E}_{Q(\theta,F)}[h(Y\vert{}X)\vert{}X]$:

$$\frac{\partial}{\partial \epsilon} \mathrm{DL}(\nu; f_\epsilon) \bigg\vert{}_{\epsilon=0} = \mathbb{E}_P \left[ \int W_0(x,y) \left( h(y\vert{}x) - \mathbb{E}_{Q(\theta,F)}[h(Y\vert{}X)\vert{}X] \right) q_0(y\vert{}x) dy \right]$$

As  in the previous derivation, this integral represents a covariance under $Q$, which allows us to shift the centering onto $W_0(x,y)$:

$$= \mathbb{E}_P \left[ \int \left( W_0(x,y) - \mathbb{E}_{Q(\theta,F)}[W_0(X,Y)\vert{}X] \right) h(y\vert{}x) q_0(y\vert{}x) dy \right]$$

Let $\Delta W(X, Y) = W_0(X, Y) - \mathbb{E}_{Q(\theta,F)}[W_0(X,Y)\vert{}X]$.

\subsubsection{Isolate the Direction   and Change the Measure}

We follow the exact same change-of-measure process used for $M(\nu; F)$ to bring the expectation back to the underwriting distribution $F$.

1. Expand $\mathbb{E}_P$ using the covariate density ratio $r(x) = \frac{dP_X(x)}{dF_X(x)}$.
2. Expand the tilted density $q_0(y\vert{}x) = w(x, y; \theta, f) f(y\vert{}x)$, where $w(x, y; \theta, f) = \frac{\exp(\theta \eta(x, y))}{Z_0(x)}$.

$$\frac{\partial}{\partial \epsilon} \mathrm{DL}(\nu; f_\epsilon) \bigg\vert{}_{\epsilon=0} = \int \int r(x) w(x, y; \theta, f) \Delta W(x, y) h(y\vert{}x) f(y\vert{}x) dy \, dF_X(x)$$

Because $f(y\vert{}x) dy \, dF_X(x) = dS(x, y)$, this is an expectation under $F$:

$$\frac{\partial}{\partial \epsilon} \mathrm{DL}(\nu; f_\epsilon) \bigg\vert{}_{\epsilon=0} = \mathbb{E}_F \left[ r(X) w(X, Y; \theta, f) \Delta W(X, Y) h(Y\vert{}X) \right]$$

\subsubsection{The Canonical Gradient}

By writing the Gâteaux derivative as an inner product $\mathbb{E}_F[\mathrm{D}_{\mathrm{DL}}(X, Y) h(Y\vert{}X)]$, we have isolated the canonical gradient  for $\mathrm{DL}(\nu; F)$:

$$\mathrm{D}_{\mathrm{DL}}(X, Y) = r(X) w(X, Y; \theta, f) \left( W_0(X, Y) - \mathbb{E}_{Q(\theta,F)}[W_0(X,Y) \mid X] \right)$$

If we expand $W_0$, the full expression for the canonical gradient is:

$$\mathrm{D}_{\mathrm{DL}}(X, Y) = r(X) w(X, Y; \theta, f) \left( \Delta \eta (\Delta \eta^\top \theta + \Delta \gamma^\top \lambda) - \left( \text{Cov}_{Q}(\eta, \eta^\top\vert{}X)\theta + \text{Cov}_{Q}(\eta, \gamma^\top\vert{}X)\lambda \right) \right)$$

This structure   matches $\mathrm{D}_M(X,Y)$: both gradients are built on the form 
\begin{equation*}
   r(X) w(X, Y) \left[ \text{Target} - \mathbb{E}_Q[\text{Target}\vert{}X] \right].  
\end{equation*}

 \subsection{Sanity Check: Conditional Mean-Zero}\label{section:conditional_mean_zero}
We show that the canonical gradient $\mathrm{D}_M(X, Y)$ is conditionally mean-zero under the baseline distribution $F$ by  explicitly evaluating the conditional expectation.
We recall the definition:

$$D_M(X, Y) = r(X) w(X, Y; \theta, f) \Big( \gamma_k(X, Y) - \mathbb{E}_{Q(\theta, F)}[\gamma_k(X, Y) \mid X] \Big).$$

We evaluate the conditional expectation of this quantity under the biased baseline distribution $F$, conditioning on $X=x$. This takes the form of an integral against the baseline conditional density $f(y \mid x)$:
\begin{align*}
&\mathbb{E}_F [D_M(X, Y) \mid X=x] 
\\& \quad = \int D_M(x, y) f(y \mid x) \, dy 
\\& \quad = \int r(x) w(x, y; \theta, f) \Big( \gamma_k(x, y) - \mathbb{E}_{Q(\theta, F)}[\gamma_k(X, Y) \mid X=x] \Big) f(y \mid x) \, dy
\\& \quad 
= r(x) \int w(x, y; \theta, f) \Big( \gamma_k(x, y) - \mathbb{E}_{Q(\theta, F)}[\gamma_k(X, Y) \mid X=x] \Big) f(y \mid x) \, dy.
\end{align*}
Next, we recognize the change of measure. The tilting weight is defined as
\begin{equation*}
w(x, y; \theta, f) = \frac{\exp(\theta \eta(x, y))}{Z_0(x)}.
\end{equation*}
The product of this weight and the baseline density $f(y \mid x)$ exactly yields the conditional density of the exponentially tilted distribution $Q(\theta, F)$, denoted as $q_0(y \mid x)$:

$$q_0(y \mid x) = w(x, y; \theta, f) f(y \mid x).$$

Substituting $q_0(y \mid x)$ into the integral   shifts the expectation from the baseline measure $F$ to the tilted measure $Q$:
\begin{align*}
&\mathbb{E}_F [D_M(X, Y) \mid X=x] 
\\&= r(x) \int \Big( \gamma_k(x, y) - \mathbb{E}_{Q(\theta, F)}[\gamma_k(X, Y) \mid X=x] \Big) q_0(y \mid x) \, dy \\
&= r(x) \left( \int \gamma_k(x, y) q_0(y \mid x) \, dy - \int \mathbb{E}_{Q(\theta, F)}[\gamma_k(X, Y) \mid X=x] q_0(y \mid x) \, dy \right) \\
&= r(x) \Big( \mathbb{E}_{Q(\theta, F)}[\gamma_k(X, Y) \mid X=x] - \mathbb{E}_{Q(\theta, F)}[\gamma_k(X, Y) \mid X=x] \int q_0(y \mid x) \, dy \Big) \\
&= r(x) \Big( \mathbb{E}_{Q(\theta, F)}[\gamma_k(X, Y) \mid X=x] - \mathbb{E}_{Q(\theta, F)}[\gamma_k(X, Y) \mid X=x] \Big) \\
&= 0.
\end{align*}
Next, we show that the conditional expectation of   $D_{DL}(X,Y)$ is zero.
The definition is recalled below:
$$D_{DL}(X, Y) = r(X) w(X, Y; \theta, f) \Big( W_0(X, Y) - \mathbb{E}_{Q(\theta, F)}[W_0(X, Y) \mid X] \Big).$$

Conditioning on $X=x$ under the baseline distribution $F$, we express the conditional expectation as an integral against the baseline conditional density $f(y \mid x)$:
\begin{align*}
&\mathbb{E}_F [D_{DL}(X, Y) \mid X=x] 
\\& \quad = \int D_{DL}(x, y) f(y \mid x) \, dy \\
&
\quad
= \int r(x) w(x, y; \theta, f) \Big( W_0(x, y) - \mathbb{E}_{Q(\theta, F)}[W_0(X, Y) \mid X=x] \Big) f(y \mid x) \, dy
\\ & \quad
=
r(x) \int w(x, y; \theta, f) \Big( W_0(x, y) - \mathbb{E}_{Q(\theta, F)}[W_0(X, Y) \mid X=x] \Big) f(y \mid x) \, dy.
\end{align*}
Since the product of the tilting weight and the baseline density is precisely the tilted density:
$$q_0(y \mid x) = w(x, y; \theta, f) f(y \mid x),$$
substituting $q_0(y \mid x)$ transforms the integral, shifting the expectation from the baseline measure $F$ to the tilted measure $Q$:
$$\mathbb{E}_F [D_{DL}(X, Y) \mid X=x] = r(x) \int \Big( W_0(x, y) - \mathbb{E}_{Q(\theta, F)}[W_0(X, Y) \mid X=x] \Big) q_0(y \mid x) \, dy.$$

Exploiting the linearity of integration, we distribute the integral across the terms. Since integrating any function against $q_0(y \mid x)$ evaluates its conditional expectation under $Q(\theta, F)$, we obtain:
\begin{align*}
&\mathbb{E}_F [D_{DL}(X, Y) \mid X=x] 
\\& \quad = r(x) \left( \int W_0(x, y) q_0(y \mid x) \, dy - \int \mathbb{E}_{Q(\theta, F)}[W_0(X, Y) \mid X=x] q_0(y \mid x) \, dy \right) \\
& \quad = r(x) \Big( \mathbb{E}_{Q(\theta, F)}[W_0(X, Y) \mid X=x] - \mathbb{E}_{Q(\theta, F)}[W_0(X, Y) \mid X=x] \int q_0(y \mid x) \, dy \Big).
\end{align*}

Because $q_0(y \mid x)$ is a normalized probability density function, the integral $\int q_0(y \mid x) \, dy$ is exactly $1$. The terms inside the parentheses   cancel:
\begin{align*}
\mathbb{E}_F [D_{DL}(X, Y) \mid X=x] &= r(x) \Big( \mathbb{E}_{Q(\theta, F)}[W_0(X, Y) \mid X=x] - \mathbb{E}_{Q(\theta, F)}[W_0(X, Y) \mid X=x] \Big) \\
&= 0.
\end{align*}

\section{Construction of the TMLE Fluctuation Submodel}\label{section:TMLE_submodel}

To construct the TMLE fluctuation submodel for a continuous density, we leverage a   property of the derived canonical gradients: they are  conditionally mean-zero under the biased observation distribution $F$. 

Recall that $q_0$ is defined in \eqref{q_zero}.
Because both $\mathrm{D}_M(X,Y)$ and $\mathrm{D}_{\mathrm{DL}}(X,Y)$ contain the tilting weight $w(X, Y; \theta, f) = \frac{q_0(Y\mid X)}{f(Y\mid X)}$, integrating them against the baseline density $f(Y\mid X)$ effectively changes the measure back to $Q$, where the centered target variables naturally sum to zero. This zero-mean property is essential for constructing a valid exponential family submodel, guaranteeing that the perturbed density remains strictly positive and properly integrates to 1.

The construction of the TMLE update proceeds as follows:

\paragraph{Step 1: Combine the Canonical Gradients}
Let $\nu = (\theta, \lambda)$ denote the $(J+K)$-dimensional target parameter vector. We concatenate the derived canonical gradients into a single joint vector function $D^\ast(X, Y)$:
\[
D^\ast(X, Y) = \begin{bmatrix} \mathrm{D}_{\mathrm{DL}}(X, Y) \\ \mathrm{D}_M(X, Y) \end{bmatrix}.
\]
Recall that each component takes the structural form 
\begin{equation*}
   r(X) w(X, Y; \theta, f) \left[ T(X, Y) - \mathbb{E}_Q[T(X, Y) \mid X] \right].   
\end{equation*}
 Consequently, their conditional expectation under the baseline distribution $F$ vanishes exactly:
\[
\mathbb{E}_F [D^\ast(X, Y) \mid X=x] = \int D^\ast(x, y) f(y\mid x) \, dy = \mathbf{0} \quad \forall x \in \mathcal{X}.
\]

\paragraph{Step 2: Construct the Exponential Fluctuation Submodel}
We introduce a $(J+K)$-dimensional fluctuation parameter $\epsilon \in \mathbb{R}^{J+K}$. The parametric extension of the continuous conditional density $f(y\mid x)$ is formulated as an exponential tilt:
\[
f_\epsilon(y\mid x) = \frac{f(y\mid x) \exp\left( \epsilon^\top D^\ast(x, y) \right)}{C(\epsilon, x)},
\]
where $C(\epsilon, x)$ is the normalizing constant (partition function) required to ensure the conditional density integrates to 1 over the outcome space:
\[
C(\epsilon, x) = \int f(z\mid x) \exp\left( \epsilon^\top D^\ast(x, z) \right) \, dz.
\]

\paragraph{Step 3: Verify the Score Matches the Canonical Gradient}
For this construction to serve as a valid TMLE submodel, its score (the gradient of the log-likelihood with respect to the fluctuation parameter) evaluated at $\epsilon = \mathbf{0}$ must precisely equal the canonical gradient $D^\ast(x, y)$.

Taking the logarithm of the submodel yields:
\[
\log f_\epsilon(y\mid x) = \log f(y\mid x) + \epsilon^\top D^\ast(x, y) - \log C(\epsilon, x).
\]
Differentiating with respect to the vector $\epsilon$ gives:
\[
\left. \nabla_\epsilon \log f_\epsilon(y\mid x) \right|_{\epsilon=\mathbf{0}} = D^\ast(x, y) - \frac{\left. \nabla_\epsilon C(\epsilon, x) \right|_{\epsilon=\mathbf{0}}}{C(\mathbf{0}, x)}.
\]
Since $C(\mathbf{0}, x) = \int f(z\mid x) \, dz = 1$, we evaluate the derivative of the integral term:
\[
\left. \nabla_\epsilon C(\epsilon, x) \right|_{\epsilon=\mathbf{0}} = \int f(z\mid x) D^\ast(x, z) \exp(0) \, dz = \mathbb{E}_F [D^\ast(X, Y) \mid X=x].
\]
As established in Step 1, this conditional expectation is exactly zero, causing the second term to vanish . Thus, the score of the fluctuation submodel exactly recovers the canonical gradient:
\[
\left. \nabla_\epsilon \log f_\epsilon(y\mid x) \right|_{\epsilon=\mathbf{0}} = D^\ast(x, y).
\]
This confirms that maximizing the submodel likelihood with respect to $\epsilon$ successfully targets the estimating equations.

\section{End-to-End Workflow and Computational Details}\label{section:end_to_end}
Our complete estimation pipeline integrates the core components (diffusion, TMLE, KL optimization, FK steering) into a unified procedure consisting of three primary phases:
\begin{enumerate}
    \item \textbf{Base Density Estimation:} We first estimate the baseline conditional density $f$ from the biased high-resolution simulation data using a continuous-time conditional diffusion model. Full architectural details for this score-based estimator are provided in Appendix \ref{section:diffusion_architecture}.
    
    \item \textbf{TMLE Debiasing:} Next, we correct the diffusion model's  regularization bias by applying Targeted Maximum Likelihood Estimation (TMLE). This targeted exponential fluctuation is executed efficiently via inference-time Feynman-Kac (FK) steering, as detailed in Appendix \ref{section:solve_mle} and   Algorithm \ref{alg:tmle_grad_ascent}.
    
    \item \textbf{Constrained Optimization:} Finally, using the debiased density, we numerically solve the core constrained KL optimization problem \eqref{constrained_optimization}, using 
    mini-batch Stochastic Gradient
Descent-Ascent (SGDA) procedure.
    To bypass the prohibitive computational burden of iteratively retraining the generative model, we once again leverage FK steering to instantiate the MNAR exponential tilts and evaluate the necessary expectations directly at inference time (detailed in Appendix \ref{section:solve_estimating_eq}).
\end{enumerate}
This procedure is summarized in Algorithm \ref{alg:end_to_end_pipeline}.
\begin{algorithm}[htbp]
\caption{End-to-End Multifidelity Emulation via Diffusion, TMLE, and FK Steering}
\label{alg:end_to_end_pipeline}
\begin{algorithmic}[1]
\REQUIRE Biased high-resolution dataset $\mathcal{D}_F = \{(X_i, Y_i)\}_{i=1}^n$, unbiased low-resolution regional moment constraints $\{\bar{\gamma}_k\}_{k=1}^K$ over partitions $\{A_k\}_{k=1}^K$, number of FK particles $M$.

\STATE \textbf{Phase 1: Base Density Estimation (Section \ref{section:diffusion_architecture})}
\STATE Pre-process $\mathcal{D}_F$: Log-transform and standardize physical targets $Y_i \to \tilde{Y}_i$ to ensure non-negativity and align with the standard Gaussian prior.
\STATE Train the Conditioned Residual MLP score network $s_\phi(\tilde{y}_t, t, c)$ on the transformed dataset using the time-dependent denoising score matching objective.
\STATE \textbf{Ensure:} A learned baseline conditional diffusion score model $s_\phi$, incorporating Classifier-Free Guidance ($\omega \approx 1.0$) for robust statistical coverage.

\vspace{0.15cm}
\STATE \textbf{Phase 2: TMLE Debiasing (Section \ref{section:solve_mle})}
\STATE \textit{Objective: Eliminate the generative model's regularization bias via an exponential fluctuation, rendering estimation locally insensitive to first-order errors.}
\STATE Invoke \textbf{Algorithm \ref{alg:tmle_grad_ascent}} (Gradient Ascent for TMLE using FK steering) using dataset $\mathcal{D}_F$ and base score $s_\phi$.
\STATE \textbf{Ensure:} Optimal targeted fluctuation parameter $\hat{\epsilon}$.

\vspace{0.15cm}
\STATE \textbf{Phase 3: Constrained KL Optimization (Section \ref{section:solve_estimating_eq})}
\STATE Invoke \textbf{Algorithm \ref{alg:sgda_estimating_eq}} (FK-Steered Primal-Dual SGDA) initialized with $s_\phi$ and $\hat{\epsilon}$, anchoring the system to the low-resolution targets $\{\bar{\gamma}_k\}_{k=1}^K$.
\STATE \textbf{Ensure:} Optimal MNAR tilting parameter $\theta^*$ and dual constraint multipliers $\lambda^*$.

\vspace{0.15cm}
\STATE \textbf{Final Generation / Inference}
\STATE To generate a calibrated, unbiased sample $Y \sim P_{Y \mid X}$ for any given covariate $X$:
\STATE \quad 1. Simulate $M$ concurrent particles via FK steering using base score $s_\phi$.
\STATE \quad 2. Use FK steering with difference potentials to target the composite terminal reward:
\[ \text{Reward}(Y) = \exp\left( \hat{\epsilon}^\top D^\ast(X, Y) + {\theta^*}^\top \eta(X, Y) \right) \]
\STATE \quad 3. Apply the inverse transformation (exponentiation/destandardization) to the surviving particles to recover physical units.
\end{algorithmic}
\end{algorithm}

\subsection{Diffusion Model Architecture}\label{section:diffusion_architecture}
We model the baseline conditional density using a continuous-time score-based diffusion model equipped with Classifier-Free Guidance (CFG).
\citep{ho_classifierfree_2021}. Because the target is a one-dimensional scalar rather than a high-dimensional image, we do not use complicated U-Net structure designed for images. Instead, we employ a lightweight Conditioned Residual Multilayer Perceptron (MLP) backbone, injecting continuous high-dimensional covariates via Adaptive Layer Normalization (AdaLN). This lightweight architecture is computationally optimal for our pipeline, as it easily fits into GPU memory and allows for the rapid, parallel evaluation of thousands of concurrent particles required by the Feynman-Kac (FK) steering mechanism at inference time.

\subsubsection{Target Transformation and the Forward SDE}
Physical simulation outputs (such as pricing volumes or catastrophe risks) frequently possess strict non-negativity constraints. The raw target variable $y$ is first log-transformed to map its support to the unconstrained real line $\mathbb{R}$, and then standardized into $\tilde{y}$. This transformation is critical; it aligns the data with the standard Gaussian prior of the diffusion process and structurally prevents the reverse SDE from generating invalid negative physical quantities.

Within the score-based framework, the forward diffusion process is governed by a continuous-time It\^o SDE acting on $\tilde{y}$:
$$ d\tilde{y} = f(\tilde{y}, t)dt + g(t)dw $$
where $w$ is the standard Wiener process. This forward process systematically perturbs the empirical data distribution into a tractable prior (e.g., $\mathcal{N}(0, 1)$) over time $t \in [0, T]$. The perturbation allows us to derive the transition kernel $p_{0t}(\tilde{y}_t \mid \tilde{y}_0)$, a Gaussian distribution whose mean and variance are determined by the drift $f(\cdot, t)$ and diffusion $g(t)$ coefficients.

\subsubsection{Denoising Score Matching Objective}
To learn the score function without requiring access to the intractable marginal distribution $p_t(\tilde{y}_t)$, we optimize the time-dependent denoising score matching objective. The network weights $\phi$ are updated by minimizing the expected squared difference between the network's output and the exact gradient of the known transition kernel:
$$ \mathcal{L}(\phi) = \mathbb{E}_{t \sim \mathcal{U}[0, T], \tilde{y}_0 \sim \mathcal{D}, \tilde{y}_t \sim p_{0t}} \left[ \lambda(t) \left\| s_\phi(\tilde{y}_t, t, c) - \nabla_{\tilde{y}_t} \log p_{0t}(\tilde{y}_t \mid \tilde{y}_0) \right\|_2^2 \right] $$
where $\lambda(t)$ is a positive weighting function that balances the loss across different noise scales.

\subsubsection{Covariate Encoder and CFG Trade-offs}
Because the input space consists  of continuous covariates, the observable vector $X$ is passed through a projection MLP to yield a unified covariate latent representation:
$$ h_X = \text{MLP}(X) \in \mathbb{R}^k $$
To implement CFG, we introduce a learnable null token $h_\emptyset \in \mathbb{R}^k$. During training, the conditioning vector $h_X$ is randomly replaced by $h_\emptyset$ according to a predefined dropout probability. During the sampling phase, the conditional and unconditional scores are linearly combined to control the variance-fidelity trade-off:
$$ \tilde{s}_\phi(\tilde{y}_t, t, c) = s_\phi(\tilde{y}_t, t, \emptyset) + \omega \cdot \Big(s_\phi(\tilde{y}_t, t, h_X) - s_\phi(\tilde{y}_t, t, \emptyset)\Big) $$
where $\omega \ge 1$ defines the guidance scale. 

\textbf{Remark on Guidance Scale:} In standard generative modeling, setting $\omega \gg 1$ increases sample fidelity by intentionally truncating the tails of the distribution. However, in statistical emulation, these tails represent genuine physical uncertainty and variance. Artificially shrinking the conditional variance of the baseline estimate $f(y \mid x)$ places an unreasonable burden on the downstream TMLE step, forcing the exponential fluctuation to reconstruct massive under-dispersion rather than simply correcting localized regularization bias. Therefore, for robust statistical estimation, we constrain $\omega \approx 1.0$, prioritizing valid distributional coverage over point-estimate fidelity.

\subsubsection{AdaLN Residual Score Network Backbone}
The core network, denoted as $s_\phi(\tilde{y}_t, t, c)$, functions as the parameterized score estimator. The continuous time step $t$ is mapped to a dense vector via sinusoidal positional encodings. The unified conditioning vector $c$ is formed by fusing this temporal embedding with the covariate latent representation (or the null token).

Conditioning is integrated into the residual MLP exclusively through AdaLN. For any given hidden layer activation $a$, the conditioning vector $c$ is projected to dynamically regress the scale parameter $\gamma(c)$ and the shift parameter $\beta(c)$:
$$ \text{AdaLN}(a, c) = \gamma(c) \cdot \text{LayerNorm}(a) + \beta(c). $$
Following a sequence of these conditioned residual blocks, a final linear layer projects the modulated hidden state back down to a single scalar in $\mathbb{R}$, directly outputting the estimated one-dimensional score.

\subsection{Solving the Maximum Likelihood Problem for TMLE via FK Steering}\label{section:solve_mle}
This section details how Feynman-Kac (FK) steering is leveraged to perform Targeted Maximum Likelihood Estimation (TMLE) on the initial diffusion-based conditional density estimate, $f$.
Appendix \ref{section:grad_asc} and Algorithm \ref{alg:tmle_grad_ascent} outlines the core gradient ascent strategy used to solve the TMLE optimization problem. Because this optimization step depends  on evaluating the canonical gradient, the explicit computational procedure for constructing this gradient is detailed in Appendix \ref{section:can_grad_eval} and Algorithm \ref{alg:eval_canonical_grad}. 
\subsubsection{Gradient Ascent Algorithm}\label{section:grad_asc}
To determine the optimal TMLE fluctuation parameter $\epsilon$, we maximize the empirical log-likelihood of the observed underwriting dataset $\mathcal{D}_F = \{(X_i,Y_i)\}_{i=1}^n$ under the exponential fluctuation submodel
\begin{equation*}
f_\epsilon(y\mid x)
=
\frac{f(y\mid x)\exp\!\left\{\epsilon^\top D^\ast(x,y)\right\}}
{C(\epsilon,x)},
\end{equation*}
where
\begin{equation*}
C(\epsilon,x)
=
\int f(y\mid x)
\exp\!\left\{\epsilon^\top D^\ast(x,y)\right\}\,dy.
\end{equation*}
Here, $D^\ast(x,y)$ denotes the joint canonical-gradient direction derived in Appendix \ref{section:proof:theorem:canonical_gradient}. We showed in
Appendix \ref{section:conditional_mean_zero} that both components of $D^\ast$ are conditionally mean-zero under the baseline   distribution $F$:
\begin{equation*}
\mathbb{E}_F[D^\ast(X,Y)\mid X]=0.
\end{equation*}
Consequently, the score of the fluctuation submodel at the baseline $\epsilon=0$ is exactly $D^\ast(X,Y)$, making this an appropriate canonical-gradient fluctuation direction for TMLE.

Because $f_\epsilon$ is an exponential-family model, its empirical log-likelihood is globally concave in $\epsilon$ (and is strictly concave under the usual nondegeneracy conditions). Hence, the fluctuation parameter can be obtained by standard gradient-based optimization. Importantly, the conditional expectations appearing in the gradient can be approximated using FK steering.

Given the observed dataset, the empirical log-likelihood is
\begin{equation*}
\mathcal{L}(\epsilon)
=
\frac{1}{n}\sum_{i=1}^n
\left[
\log f(Y_i\mid X_i)
+
\epsilon^\top D^\ast(X_i,Y_i)
-
\log C(\epsilon,X_i)
\right].
\end{equation*}
During the fluctuation step, the baseline diffusion density $f(Y_i\mid X_i)$ is fixed. Therefore, differentiating with respect to $\epsilon$ gives
\begin{equation*}
\nabla_\epsilon\mathcal{L}(\epsilon)
=
\frac{1}{n}\sum_{i=1}^n
\left[
D^\ast(X_i,Y_i)
-
\nabla_\epsilon\log C(\epsilon,X_i)
\right].
\end{equation*}
By the standard exponential-family identity,
\begin{equation*}
\nabla_\epsilon\log C(\epsilon,x)
=
\mathbb{E}_{f_\epsilon}
\left[
D^\ast(x,Y)\mid X=x
\right].
\end{equation*}
Thus,
\begin{equation*}
\boxed{
\nabla_\epsilon\mathcal{L}(\epsilon)
=
\frac{1}{n}\sum_{i=1}^n
\left[
D^\ast(X_i,Y_i)
-
\mathbb{E}_{f_\epsilon}
\left[
D^\ast(X_i,Y)\mid X_i
\right]
\right].
}
\end{equation*}
Maximum likelihood estimation of $\epsilon$ using gradient ascent proceeds as follows.
Let $\epsilon^{(t)}$ denote the fluctuation parameter at iteration $t$. The gradient-ascent procedure proceeds as follows:

\begin{enumerate}
    \item \textbf{Sample via FK steering.}
    For each observed covariate $X_i$, invoke FK steering on the baseline diffusion model $s_\phi$ using the current TMLE tilting potential
    \begin{equation*}
    R_{\epsilon^{(t)}}(x,y)
    =
    \exp\!\left\{
    (\epsilon^{(t)})^\top D^\ast(x,y)
    \right\}.
    \end{equation*}
    This produces approximate samples from
    $f_{\epsilon^{(t)}}(\cdot\mid X_i)$.

    \item \textbf{Extract particles.}
    Denote the resulting FK-steered particles by
    \begin{equation*}
    \left\{
    Y_{m,i}^{(f_\epsilon)}
    \right\}_{m=1}^M.
    \end{equation*}
    These particles are approximately distributed according to
    $f_{\epsilon^{(t)}}(\cdot\mid X_i)$.

    \item \textbf{Approximate the conditional expectation.}
    The conditional expectation in the likelihood gradient is estimated by the Monte Carlo average
    \begin{equation*}
    \widehat{
    \mathbb{E}_{f_{\epsilon^{(t)}}}
    \left[
    D^\ast(X_i,Y)\mid X_i
    \right]
    }
    =
    \frac{1}{M}
    \sum_{m=1}^M
    D^\ast
    \left(
    X_i,Y_{m,i}^{(f_\epsilon)}
    \right).
    \end{equation*}

    \item \textbf{Compute the empirical gradient.}
    Substituting the Monte Carlo approximation gives
    \begin{equation*}
    \widehat{\nabla_\epsilon\mathcal{L}}
    \left(\epsilon^{(t)}\right)
    =
    \frac{1}{n}
    \sum_{i=1}^n
    \left[
    D^\ast(X_i,Y_i)
    -
    \frac{1}{M}
    \sum_{m=1}^M
    D^\ast
    \left(
    X_i,Y_{m,i}^{(f_\epsilon)}
    \right)
    \right].
    \end{equation*}

    \item \textbf{Update the fluctuation parameter.}
    Using a step size $\alpha_t>0$, perform
    \begin{equation*}
    \epsilon^{(t+1)}
    =
    \epsilon^{(t)}
    +
    \alpha_t
    \widehat{\nabla_\epsilon\mathcal{L}}
    \left(\epsilon^{(t)}\right).
    \end{equation*}
\end{enumerate}
The procedure is repeated until the empirical gradient computed in Step 4 is sufficiently close to zero, yielding the targeted fluctuation parameter $\hat{\epsilon}$.

\subsubsection{Numerical Evaluation of Canonical Gradient}\label{section:can_grad_eval}
The above procedure requires evaluation of the canonical gradient.
We recall that the  closed-form components of the canonical gradient are given by:
\begin{align*}
\mathrm{D}_{M_k}(X, Y) &= r(X) w(X, Y; \theta, f) \Big( \gamma_k(X, Y) - \mathbb{E}_{Q(\theta, F)}[\gamma_k(X, Y) \mid X] \Big),
k=1, \ldots, K
\\
\mathrm{D}_{\mathrm{DL}}(X, Y) &= r(X) w(X, Y; \theta, f) \Big( W_0(X, Y) - \mathbb{E}_{Q(\theta, F)}[W_0(X, Y) \mid X] \Big),
\end{align*}
Evaluating these components presents varying degrees of computational difficulty. Because the covariate density ratio $r(X) = dP_X(X) / dF_X(X)$ and the moment functions $\gamma_k(X, Y)$ are analytically known, they are computationally trivial to evaluate. Similarly, the dual target function $W_0(X, Y) = \Delta \eta(X, Y) \left( \Delta \eta(X, Y)^\top \theta + \Delta \gamma_k(X, Y)^\top \lambda \right)$, where $\Delta$ denotes a variable centered by its conditional expectation under $Q(\theta, F)$, can be efficiently computed. This function, along with any other expectations taken with respect to $Q(\theta, F)$, is easily approximated using empirical Monte Carlo averages from  tilted samples
generated by FK steering.

The primary computational bottleneck lies in evaluating the normalized exponential tilting weight, $w(X, Y; \theta, f) = \exp(\theta \eta(X, Y)) / Z_0(X)$. This term requires estimating the intractable partition function $Z_0(X) = \int \exp(\theta \eta(X, z)) f(z \mid X) \, dz$, a challenge we address in the subsequent discussion.

Evaluating the tilting weight $w(X, Y; \theta, f)$ requires computing the intractable normalizing constant $Z_0(X) = \int \exp(\theta \eta(X, z)) f(z \mid X) \, dz$ in its denominator. Fortunately, this partition function can be efficiently estimated as a natural mathematical byproduct of Feynman-Kac (FK) steering.

Instead of applying an exponential tilt solely at the end of the generative process, FK steering distributes the tilting across the $K$ time steps of the reverse diffusion trajectory (from pure noise at $t_K$ to generated data at $t_0$). It defines a sequence of intermediate potential functions $G_k(Y_{t_k})$ such that their product   recovers the target exponential tilt: 
$$\prod_{k=1}^K G_k(Y_{t_k}) \approx \exp(\theta \eta(X, Y_{t_0})).$$

This is operationalized via a Sequential Monte Carlo particle filter that simulates $M$ concurrent diffusion trajectories. At each step $k$, the algorithm executes three operations: mutation, which advances the $M$ particles using the base score $s_\phi$; weighting, which evaluates the intermediate potential to assign each particle an unnormalized weight $W_k^{(m)} = G_k(Y_{t_{k-1}}^{(m)})$; and resampling, which duplicates particles in high-reward regions and prunes those in low-reward regions based on their proportional weights. 

By time $t_0$, the surviving particles $\{Y_{t_0}^{(m)}\}_{m=1}^M$ are approximately distributed according to the tilted measure $Q(\theta, F \mid X)$, directly supplying the samples required to evaluate the expectations $\mathbb{E}_Q[\cdot]$ in the estimating equations. This particle filter simultaneously yields an unbiased estimator of the intractable normalizing constant $Z_0(X)$. By computing the average unnormalized weight of the particles at each step, 
$$\bar{W}_k = \frac{1}{M} \sum_{m=1}^M W_k^{(m)},$$
the partition function is efficiently estimated as the product of these step-wise averages: 
$$\hat{Z}_0(X) = \prod_{k=1}^K \bar{W}_k.$$
\begin{algorithm}[t]
\caption{Gradient Ascent for TMLE Fluctuation Parameter}
\label{alg:tmle_grad_ascent}
\begin{algorithmic}[1]
\REQUIRE Observed dataset $\mathcal{D}_F = \{(X_i, Y_i)\}_{i=1}^n$, learned base score $s_\phi$, estimating equation parameters $\nu = (\theta, \lambda)$, number of particles $M$, learning rate sequence $\{\alpha_t\}$, convergence tolerance $\tau$.

\STATE \textbf{Phase 1: Baseline Pre-computation}
\FOR{each observation $(X_i, Y_i) \in \mathcal{D}_F$}
    \STATE Evaluate the fixed baseline canonical gradient $D^\ast(X_i, Y_i)$ by invoking \textbf{Algorithm \ref{alg:eval_canonical_grad}}.
\ENDFOR

\STATE \textbf{Phase 2: TMLE Optimization Loop}
\STATE Initialize fluctuation parameter $\epsilon^{(0)} = \mathbf{0}$ and iteration counter $t = 0$.
\REPEAT
    \FOR{each covariate $X_i \in \mathcal{D}_F$}
        \STATE \textbf{Step 1: Sample via FK Steering}
        \STATE Invoke FK steering on the base diffusion model $s_\phi$ using $M$ trajectories, targeting the TMLE submodel reward:
        \begin{equation*}
            \text{Reward}(y) = \exp\!\left\{ (\epsilon^{(t)})^\top D^\ast(X_i, y) \right\}
        \end{equation*}
        \STATE \textit{(Note: Evaluating this reward during generation dynamically invokes Algorithm \ref{alg:eval_canonical_grad} for intermediate particles $y$).}
        
        \STATE \textbf{Step 2: Extract Particles}
        \STATE Extract the generated particles $\{Y_{m,i}^{(f_\epsilon)}\}_{m=1}^M$, which are approximately distributed as $f_{\epsilon^{(t)}}(\cdot \mid X_i)$.
        
        \STATE \textbf{Step 3: Approximate Conditional Expectation}
        \STATE Evaluate $D^\ast(X_i, Y_{m,i}^{(f_\epsilon)})$ for each particle via \textbf{Algorithm \ref{alg:eval_canonical_grad}}.
        \STATE Compute the Monte Carlo average:
        \begin{equation*}
            \widehat{\mathbb{E}}_{f_{\epsilon^{(t)}}} \left[ D^\ast(X_i,Y) \mid X_i \right] = \frac{1}{M} \sum_{m=1}^M D^\ast\left(X_i, Y_{m,i}^{(f_\epsilon)}\right)
        \end{equation*}
    \ENDFOR
    
    \STATE \textbf{Step 4: Compute the Empirical Gradient}
    \STATE Aggregate the errors across the full dataset:
    \begin{equation*}
        \widehat{\nabla_\epsilon\mathcal{L}} \left(\epsilon^{(t)}\right) = \frac{1}{n} \sum_{i=1}^n \left[ D^\ast(X_i, Y_i) - \widehat{\mathbb{E}}_{f_{\epsilon^{(t)}}} \left[ D^\ast(X_i,Y) \mid X_i \right] \right]
    \end{equation*}
    
    \STATE \textbf{Step 5: Update Fluctuation Parameter}
    \STATE Apply the gradient ascent step:
    \begin{equation*}
        \epsilon^{(t+1)} = \epsilon^{(t)} + \alpha_t \widehat{\nabla_\epsilon\mathcal{L}} \left(\epsilon^{(t)}\right)
    \end{equation*}
    
    \STATE $t \leftarrow t + 1$
\UNTIL{$\left\| \widehat{\nabla_\epsilon\mathcal{L}} \left(\epsilon^{(t-1)}\right) \right\| < \tau$}

\ENSURE The optimal targeted fluctuation parameter $\hat{\epsilon} = \epsilon^{(t)}$.
\end{algorithmic}
\end{algorithm}

\begin{algorithm}[t]
\caption{Evaluate the Canonical Gradient $D^\ast(X, Y)$}
\label{alg:eval_canonical_grad}
\begin{algorithmic}[1]
\REQUIRE Evaluation point $(X, Y)$, parameters $\nu = (\theta, \lambda)$, learned base score $s_\phi$, known density ratio $r(X)$,   partitions $\{A_k\}_{k=1}^K$, number of particles $M$.

\STATE \textbf{Step 1: Exponential Tilting via Feynman-Kac (FK) Steering}
\STATE Invoke FK steering on the base diffusion model $s_\phi$ with the terminal reward function $\exp(\theta \eta(X, y))$ using $M$ trajectories.
\STATE Extract the generated tilted samples $\{Y^{(Q)}_{m}\}_{m=1}^M \sim Q(\theta, F \mid X)$ and the byproduct normalizing constant estimate $\hat{Z}_0(X) = \prod_{k=1}^K \bar{W}_k$.
\STATE Evaluate the exact tilting weight for the fixed evaluation point $(X, Y)$: 
\[ w(X, Y; \theta, f) = \frac{\exp\left(\theta \eta(X, Y)\right)}{\hat{Z}_0(X)} \]

\STATE \textbf{Step 2: Base Moments under $Q$}
\STATE Compute empirical base moments under $Q$ using the extracted particles $\{Y^{(Q)}_{m}\}_{m=1}^M$:
\[ \hat{\mu}_Y, \quad \bar{\eta}, \quad \widehat{\operatorname{Cov}}[\eta, \eta], \quad \text{and} \quad \widehat{\operatorname{Cov}}[\eta, Y]. \]

\STATE \textbf{Step 3: Compute $D_M$ Components}
\FOR{$k = 1$ \TO $K$}
    \STATE $D_{M_k}(X, Y) = r(X) w(X, Y; \theta, f)\, \mathbb{I}(X \in A_k) \left(Y - \hat{\mu}_Y\right)$
\ENDFOR

\STATE \textbf{Step 4: Compute $D_{DL}$ Component}
\STATE Compute centered variables for the evaluation point: $\Delta \eta = \eta(X, Y) - \bar{\eta}$ and $\Delta Y = Y - \hat{\mu}_Y$.
\STATE Evaluate the dual parameter target function:
\[ W_0(X, Y) = \Delta \eta \left( \Delta \eta^\top \theta + \sum_{k=1}^K \mathbb{I}(X \in A_k) \Delta Y \lambda_k \right) \]
\STATE Compute its conditional expectation under $Q$ using the pre-computed empirical covariances:
\[ \hat{\mathbb{E}}_Q[W_0 \mid X] = \widehat{\operatorname{Cov}}[\eta, \eta]\, \theta + \sum_{k=1}^K \mathbb{I}(X \in A_k) \widehat{\operatorname{Cov}}[\eta, Y]\, \lambda_k \]
\STATE Evaluate the estimating equation gradient component:
\[ D_{DL}(X, Y) = r(X) w(X, Y; \theta, f) \left( W_0(X, Y) - \hat{\mathbb{E}}_Q[W_0 \mid X] \right) \]

\STATE \textbf{Return:} Canonical gradient vector $D^\ast(X, Y) = \left[ D_{M_1}, \dots, D_{M_K}, D_{DL} \right]^\top$
\end{algorithmic}
\end{algorithm}

\subsection{Solving the Core Optimization Problem via FK-Steered SGDA}\label{section:solve_estimating_eq}
A critical computational challenge in solving the estimating equations \eqref{estimating_equation_1} and \eqref{estimating_equation_2} is computing the conditional expectations and covariances under the exponentially tilted measure $Q(\theta, F)$. Because this system of equations derives from the first-order conditions of a constrained optimization problem, identifying a valid root corresponds to finding a saddle point of the underlying Lagrangian. We achieve this using a mini-batch Stochastic Gradient Descent-Ascent (SGDA) procedure (Algorithm \ref{alg:sgda_estimating_eq}) that   integrates Feynman-Kac (FK) steering to evaluate the necessary moments.

Our low-resolution moment constraints are defined over partition $A_k \subset \mathcal{X}$, taking the explicit form $\gamma_k(X,Y) = Y\,\mathbb{I}(X\in A_k)$. Because the partition indicator $\mathbb{I}(X\in A_k)$ is purely deterministic given the covariates $X$, it factors completely outside of any conditional expectation over $Y$. Consequently, the required terms simplify to:
\begin{equation*}
    \mathbb{E}_{Q(\theta,F)}[\gamma_k(X, Y) \mid X] = \mathbb{I}(X \in A_k) \mathbb{E}_{Q(\theta,F)}[Y \mid X]
\end{equation*}
and for the covariance required by the primal gradient:
\begin{equation*}
    \operatorname{Cov}_{Q(\theta,F)}[\eta(X,Y), \gamma_k(X, Y) \mid X] = \mathbb{I}(X \in A_k) \operatorname{Cov}_{Q(\theta,F)}[\eta(X,Y), Y \mid X].
\end{equation*}
This factorization reveals that we only need to compute the base conditional moments of $Y$ and $\eta(X,Y)$. 

\textbf{Step 1: Feynman-Kac Steering Sampling.} At each optimization iteration $t$, we decouple the optimization step from the generation process. For a mini-batch of covariates, we must draw samples from the tilted distribution $Q(\theta^{(t)}, F \mid X_i)$. We leverage  FK steering to do it without re-training.

The baseline diffusion density $f$ we are tilting has already been corrected by TMLE to eliminate regularization bias. However, this TMLE correction is not the result of  network retraining; rather, it is itself an inference-time FK steering operation. Practically, the TMLE phase simply yields the saved optimal fluctuation vector $\hat{\epsilon}$. 

Therefore, letting $s_\phi$ denote the original, uncorrected baseline score, we simulate $M$ concurrent particles. At each intermediate diffusion step, particles are mutated using $s_\phi$ and dynamically resampled according to intermediate potentials. To simultaneously apply the TMLE nuisance correction and the MNAR exponential tilt in a single unified step, these potentials target the composite terminal reward:
\begin{equation*}
    \text{Reward}(y) = \exp\left( \hat{\epsilon}^\top D^\ast(X_i, y) + {\theta^{(t)}}^\top \eta(X_i, y) \right).
\end{equation*}
This procedure directly outputs $M$ independent synthetic Monte Carlo samples from the debiased $Q(\theta^{(t)}, F \mid X_i)$ distribution without ever requiring gradients of $\eta$, gradients of $D^\ast$, or the intractable partition function.

\textbf{Step 2: Base Moments Computation.} The $M$ surviving synthetic particles are aggregated into conditional sample means and covariances locally for each $X_i$. By computing these sufficient statistics immediately, we circumvent the need to retain high-dimensional trajectory data in memory.

\textbf{Step 3: Evaluate Stochastic Gradients.} We construct the stochastic gradients for both the objective function and the constraints. Using the known covariate density ratio $r(X)$ to change the measure from the target population to the biased baseline distribution, we multiply the unit-level base moments by the regional indicator masks $\mathbb{I}(X_i\in A_k)$. 

This formulation presents a  computational advantage: the FK particle filter is simulated only once per optimization step to generate a single synthetic cohort. This exact same cohort is reused to evaluate every partition moment constraint and its corresponding covariance simply by applying different binary indicator masks. Furthermore, the Monte Carlo variance introduced by the finite particle size $M$ naturally substitutes for the stochastic noise  in standard SGD, making the approximation theoretically sound.

\textbf{Step 4: Primal-Dual SGDA Update.} The parameter vectors are updated via alternating gradient steps. We apply a gradient descent step on the primal variable $\theta$ with learning rate $\alpha_t$ to minimize the KL divergence objective. Conversely, we apply a gradient ascent step on the dual variable $\lambda$ with learning rate $\beta_t$ to penalize constraint violations, thereby enforcing the low-resolution moments. This alternating min-max dynamic continues until the residuals of the estimating equations fall below a specified tolerance.

\begin{algorithm}[t]
\caption{Solve Estimating Equations via FK-Steered Primal-Dual SGDA}
\label{alg:sgda_estimating_eq}
\begin{algorithmic}[1]
\REQUIRE Dataset $\mathcal{D}_F$, learned base score $s_\phi$, optimal TMLE fluctuation parameter $\hat{\epsilon}$, canonical gradient function $D^\ast$, density ratio $r(X)$, partitions $\{A_k\}_{k=1}^K$, particle count $M$, initial guess $(\theta^{(0)}, \lambda^{(0)})$, learning rate sequences $\{\alpha_t\}, \{\beta_t\}$.

\STATE Initialize optimization iteration counter $t \leftarrow 0$.
\REPEAT
    \STATE Sample a mini-batch $\mathcal{B} \subset \mathcal{D}_F$ of size $B$.
    
    \STATE \textbf{Step 1: Feynman-Kac Steering Sampling}
    \FOR{each $X_i \in \mathcal{B}$}
        \STATE Invoke FK steering on the base diffusion model $s_\phi$ using $M$ particles, targeting the composite terminal reward:
        \begin{equation*}
            \text{Reward}(Y) = \exp\left( \hat{\epsilon}^\top D^\ast(X_i, Y) + {\theta^{(t)}}^\top \eta(X_i, Y) \right)
        \end{equation*}
        \STATE Extract the surviving particles to form the Monte Carlo cohort $\{Y_{i,m}\}_{m=1}^M \sim Q(\theta^{(t)}, F \mid X_i)$.
    \ENDFOR

    \STATE \textbf{Step 2: Base Moments Computation}
    \FOR{each $X_i \in \mathcal{B}$}
        \STATE Compute sample means: $\hat{\mu}_{Y,i} = \frac{1}{M} \sum_{m=1}^M Y_{i,m}$ and $\bar{\eta}_i = \frac{1}{M} \sum_{m=1}^M \eta(X_i, Y_{i,m})$.
        \STATE Compute empirical base conditional covariances:
        \begin{equation*}
            \widehat{\operatorname{Cov}}_{i}[\eta, \eta] = \frac{1}{M-1} \sum_{m=1}^M (\eta(X_i, Y_{i,m}) - \bar{\eta}_i)(\eta(X_i, Y_{i,m}) - \bar{\eta}_i)^\top
        \end{equation*}
        \begin{equation*}
            \widehat{\operatorname{Cov}}_{i}[\eta, Y] = \frac{1}{M-1} \sum_{m=1}^M (\eta(X_i, Y_{i,m}) - \bar{\eta}_i)(Y_{i,m} - \hat{\mu}_{Y,i})^\top
        \end{equation*}
    \ENDFOR

    \STATE \textbf{Step 3: Evaluate Stochastic Gradients}
    \STATE Compute the dual constraint gradient vector $\hat{M}(\theta^{(t)}, \lambda^{(t)})$ with the $k$-th element:
    \begin{equation*}
        \hat{M}_k(\theta^{(t)}, \lambda^{(t)}) = \frac{1}{B} \sum_{X_i \in \mathcal{B}} r(X_i)\, \mathbb{I}(X_i \in A_k)\, \hat{\mu}_{Y,i} - \bar{\gamma}_{k}
    \end{equation*}
    \STATE Compute the primal objective gradient:
    \begin{equation*}
        \widehat{DL}(\theta^{(t)}, \lambda^{(t)}) = \frac{1}{B} \sum_{X_i \in \mathcal{B}} r(X_i) \left( \widehat{\operatorname{Cov}}_{i}[\eta, \eta]\, \theta^{(t)} + \sum_{k=1}^K \mathbb{I}(X_i \in A_k) \widehat{\operatorname{Cov}}_{i}[\eta, Y]\, \lambda^{(t)}_k \right)
    \end{equation*}

    \STATE \textbf{Step 4: Primal-Dual SGDA Update}
    \STATE Primal Descent (minimizing the KL divergence objective):
    \begin{equation*}
        \theta^{(t+1)} = \theta^{(t)} - \alpha_t \widehat{DL}(\theta^{(t)}, \lambda^{(t)})
    \end{equation*}
    \STATE Dual Ascent (maximizing the constraint enforcement):
    \begin{equation*}
        \lambda^{(t+1)} = \lambda^{(t)} + \beta_t \hat{M}(\theta^{(t)}, \lambda^{(t)})
    \end{equation*}
    
    \STATE $t \leftarrow t + 1$
\UNTIL{convergence criteria are met}

\ENSURE Optimal parameters $(\theta^*, \lambda^*) = (\theta^{(t)}, \lambda^{(t)})$.
\end{algorithmic}
\end{algorithm}

\section{Details on Numerical Experiments}\label{section:numerical}

All three experimental scenarios share a unified data-generating framework. The exogenous covariates are drawn independently from a standard normal distribution, $X_i \sim \mathcal{N}(0,1)$. The true outcome $Y_{\mathrm{true}}$ is generated via a deterministic physical mapping that serves as the oracle, while the biased high-resolution simulation $Y_{\mathrm{bias}}$ is produced by a structurally misspecified simulator corrupted by additive Gaussian noise. During the training of the baseline diffusion model, covariate batches are drawn dynamically; however, the final constraint-matching and evaluation phase operates on a fixed cohort of $N=100$ samples.

 We use $K=8$ low-resolution moment constraints. The partitions $A_1, \dots, A_8$ are constructed dynamically by  discretizing the covariate space into eight equal-width bins using uniform intervals established between the realized minimum and maximum covariate values ($X_{\min}$ to $X_{\max}$).
For each   partition $A_k$, the corresponding low-resolution moment target $\bar{\gamma}_k$ is defined as the empirical mean of the true oracle outcome for all sample points falling within that specific bin:
\begin{equation*}
\bar{\gamma}_k = \frac{1}{\left| \{i : X_i \in A_k\} \right|} \sum_{i : X_i \in A_k} Y_{\mathrm{true}}(X_i).
\end{equation*}

\subsection{Toy Gas Compressibility Model}

In this experimental scenario, the covariate $X$ represents a pressure-like input, and the outcome $Y$ denotes a volumetric response. The true underlying physics follows a nonlinear compressibility curve, serving as a mathematically tractable proxy for a Van der Waals real gas. Conversely, the biased high-resolution simulator misspecifies this relationship by assuming a strictly linear ideal-gas law subjected to additive Gaussian noise.

\subsubsection{True Dynamics}
\[
Y_{\mathrm{true}}(X) = -4X + V_0 + 3\,\mathrm{softplus}(2X), \qquad V_0 = 20,
\]
where $\mathrm{softplus}(z) = \log(1+e^z)$. 

This synthetic oracle function is constructed to explicitly isolate physical modeling errors. The linear component, $-4X + V_0$, mimics the standard, simplistic volume-pressure relationship of an ideal gas. The softplus component introduces a smooth, nonlinear bend, which acts as a mathematically convenient stand-in for the nonlinear compressibility of a real gas. While not the exact cubic solution of the true Van der Waals equation of state, this function   captures the structural deviation from ideal gas behavior, providing a rigorous test for the algorithm's ability to correct functional misspecification.

\subsubsection{Biased High-Resolution Simulator}
\[
Y_{\mathrm{bias}} = -3 X + (V_0 - 1) + \varepsilon, \qquad \varepsilon \sim \mathcal{N}(0,1).
\]

\subsubsection{Source of Bias} The biased simulator enforces a strictly linear dependence on $X$ (the ideal gas approximation). It structurally fails to capture the nonlinear compressibility (the softplus bend)  to the true data-generating process, while also suffering from a miscalibrated slope and baseline intercept.
\subsection{First-Order Compartment Relaxation Model}

In this scenario, we simulate the temporal evolution of a well-mixed physical compartment whose state $Y(t)$ relaxes toward an input-dependent equilibrium $\mu(X)$. While the underlying ordinary differential equation (ODE) is structurally linear with respect to the state variable $Y$ (defining a standard first-order relaxation process), the target equilibrium $\mu(X)$ exhibits complex, nonlinear dependence on the exogenous covariate $X$. Both the oracle and the biased simulator share the same foundational kinetic structure, but they differ fundamentally in their equilibrium targets and kinetic rates.

\subsubsection{Shared Kinetic Dynamics}
\[
\frac{dY}{dt} = \gamma\bigl(\mu(X) - Y\bigr), \qquad Y(0)=V_0.
\]
For both models, the continuous-time system is integrated numerically using the explicit Euler method ($80$ steps with a step size of $dt=0.1$). The final observable outcome is the terminal state $Y$ at the end of the integration period.

\subsubsection{True Dynamics}
Assuming an initial state $V_0=8$ and a true relaxation rate $\gamma_{\mathrm{true}}=0.45$, the true equilibrium state is governed by a compound nonlinear function capturing both saturation and higher-order scaling:
\[
\mu_{\mathrm{true}}(X) = V_0 + 5.5\,\tanh(1.4X) + 1.8X^2.
\]

\subsubsection{Biased High-Resolution Simulator}
The misspecified simulator operates with an artificially fast relaxation rate ($\gamma_{\mathrm{bias}}=0.90$) and assumes a strictly linear equilibrium target:
\begin{align*}
\mu_{\mathrm{bias}}(X) &= -2.2X + (V_0 + 1.5), \\
Y_{\mathrm{bias}} &= Y_{\mathrm{ODE}}(\mu_{\mathrm{bias}}, \gamma_{\mathrm{bias}}) + \varepsilon, \qquad \varepsilon \sim \mathcal{N}(0, 0.75^2).
\end{align*}

\subsubsection{Source of Bias} The bias in this model is two-fold. First, it suffers from structural misspecification: the biased simulator targets a strictly linear equilibrium,  omitting the saturation dynamics ($\tanh$) and quadratic acceleration ($X^2$) of the true physical process. Second, it suffers from parametric kinetic error, enforcing a relaxation rate ($\gamma$) that is twice as fast as the ground truth, fundamentally altering the transient trajectory of the state before the terminal observation.
\subsection{Surface Adsorption Isotherm Model}

In our final scenario, we model the physical accumulation of gas molecules onto a solid surface (adsorption). The exogenous covariate $X$ represents a normalized log-pressure, which is transformed into a strictly positive effective partial pressure $p(X)$. The true underlying physics are governed by a Langmuir isotherm, a model that naturally accounts for the finite number of available binding sites. In contrast, the biased high-resolution simulator incorrectly employs Henry's Law, a rudimentary linear approximation that strictly holds only in the low-pressure limit.

\subsubsection{Effective Partial Pressure}
To enforce non-negativity, the raw covariate is mapped to physical pressure via the softplus function:
\[
p(X) = \mathrm{softplus}(X) = \log(1+e^X).
\]

\subsubsection{True Dynamics}
Assuming a baseline environmental adsorption level $Y_{\mathrm{base}}=10$, a maximum saturation capacity $q_{\mathrm{sat}}=18$, and an equilibrium binding constant $K=3$, the true adsorbed volume follows the Langmuir saturation model:
\[
Y_{\mathrm{true}}(X) = Y_{\mathrm{base}} + q_{\mathrm{sat}}\frac{K p(X)}{1+K p(X)}.
\]

\subsubsection{Biased High-Resolution Simulator}
The misspecified simulator operates under the assumption of Henry's Law with a proportionality constant (Henry's volatility) $H=8$. It also suffers from a constant zero-order calibration shift and is corrupted by additive noise ($\sigma=0.65$):
\[
Y_{\mathrm{bias}} = Y_{\mathrm{base}} - 3 + H p(X) + \varepsilon, \qquad \varepsilon \sim \mathcal{N}(0, \sigma^2).
\]

\subsubsection{Source of Bias} The structural misspecification in this model is rooted in asymptotic physical constraints. Henry's Law asserts a strictly linear relationship between partial pressure and adsorption, which theoretically implies an infinite number of binding sites. Consequently, it fails   at higher pressures by predicting unbounded adsorption growth,  missing the asymptotic saturation enforced by the true Langmuir formulation. Furthermore, the biased simulator introduces a persistent zero-order calibration error (the $-3$ baseline shift) that uniformly displaces the entire response curve.
\begin{table}[htbp]
\centering
\caption{Summary of the physical data-generating processes and structural misspecifications across the three experimental scenarios.}
\label{tab:experiments_comparison_detailed}
\begin{tabular}{@{} l >{\raggedright\arraybackslash}p{3.8cm} >{\raggedright\arraybackslash}p{4cm} >{\raggedright\arraybackslash}p{3.8cm} @{}}
\toprule
& \textbf{Gas Compressibility} & \textbf{Compartment Relaxation} & \textbf{Surface Adsorption} \\
\midrule
\textbf{Target Outcome ($Y$)} & Volumetric response & Terminal compartment state & Adsorbed volume \\
\addlinespace
\textbf{True Physics (Oracle)} & Nonlinear compressibility curve & ODE with nonlinear equilibrium $\mu(X)$ & Langmuir isotherm (physical saturation) \\
\addlinespace
\textbf{Biased Simulator} & Linear ideal-gas law + noise & ODE with linear $\mu(X)$ and fast rate $\gamma$ + noise & Henry's Law (linear in $p$) + noise \\
\addlinespace
\textbf{Nature of Bias} & Omits nonlinear compressibility bend & Misspecified equilibrium target and kinetic rate & Omits   saturation; calibration shift \\
\bottomrule
\end{tabular}
\end{table}

\subsection{Details on FK Steering Implementation}\label{section:fk}
We use python and pytorch to implement
FK steering \citep{singhalGeneralFrameworkInferencetime2025}. FK steering has many hyperparameters. We list our choices here.
\begin{enumerate}
    \item \textbf{Proposal $\tau$ (Particle Mutation):}
    While the original paper proposes using either the base transition $p_\theta$ or gradient-tilted proposals, we rely solely on the base reverse diffusion process mutated with Classifier-Free Guidance (CFG). We do not use reward-gradient guidance on the proposal. We use CFG scale $\omega=2$.

    \item \textbf{Potentials $G_t$:}
    The paper introduces DIFFERENCE, MAX, and SUM potentials that telescope to $\exp(\lambda r(x_0))$. We implemented all three but chose to use the difference formulation: $G \propto \exp(\lambda(r_\phi(t) - r_\phi(t+1)))$.

    \item \textbf{Intermediate Reward $r_\phi(x_t)$:}
    The paper offers options including Tweedie denoised $\hat{x}_0$, Monte Carlo sampling over $x_0 \sim p_\theta(\cdot \mid x_t)$, or a learned $r_\phi$. We default to the Tweedie estimator (\texttt{fk\_reward\_samples=1}): $r_\phi(y_t) = r(\hat{y}_0)$. The score model predicts $\hat{y}_0$, and we evaluate the log-potential on the physical $Y$ scale after inverse transformation. A Monte Carlo average over perturbed $y_t$ is optional, and we omit learned intermediate reward networks entirely.

    \item \textbf{Resampling Schedule:}
    The paper resamples on a discrete schedule $R$ (e.g., intervals of 20 steps) and sets $G_t=1$ elsewhere. We use a configurable interval resampling strategy (\texttt{fk\_resample\_interval}). In production, this defaults to every 20 steps (plus $t=0$). In demonstrations, we aggressively resample at every step (\texttt{interval=0}). Resampling is performed via multinomial sampling with replacement on normalized particle weights.

    \item \textbf{Tilt Strength $\lambda$:}
    Image-based experiments in \cite{singhalGeneralFrameworkInferencetime2025} often require large scaling values (e.g., $\lambda=10$). We default to $\lambda=1.0$, applied to the log-potential increments. Thus our raw reward evaluates exactly to $\theta^\top\eta$ or the composite $\theta^\top\eta + \epsilon^\top D^\ast$.

    \item \textbf{Particle Count $M$:}
    While \cite{singhalGeneralFrameworkInferencetime2025} uses small particle sets ($2$ to $16$) for image generation, we simulate $M$ particles per covariate $X$ ($M=100$ in demos, up to $500$ in full configurations). Our particle cloud is conditional on each specific $X_i$, rather than forming one global set.

    \item \textbf{Output and Partition Function:}
    \cite{singhalGeneralFrameworkInferencetime2025}'s algorithm returns the particle set $\{x_0^{(i)}\}$, often discarding all but the highest-reward particle. Instead, we retain all $M$ particles per $X_i$ to compute empirical moments and SGDA gradients. We also accumulate the partition function estimate $\hat{Z}_0(X)$ as the product of the step-wise mean incremental weights (a requirement for evaluating $D^\ast$ and TMLE that is not emphasized in the original algorithm).

\end{enumerate}
 
\end{document}